\documentclass[11pt,a4paper]{article}

\usepackage{amsthm,amsmath}
\usepackage{amsfonts}
\usepackage{times}
\usepackage{graphicx}
\usepackage{color}{\tiny }
\usepackage{multirow}
\usepackage{hyperref}
\usepackage{color,soul}
\usepackage{xr}
\usepackage{rotating}
\usepackage{bbm}
\usepackage{latexsym}

\usepackage[justification=justified,singlelinecheck=false,font=small,labelfont=bf]{caption}
\usepackage{subcaption}
\usepackage{import}

\newcommand{\captionfonts}{\normalsize}

\makeatletter  
\long\def\@makecaption#1#2{%
  \vskip\abovecaptionskip
  \sbox\@tempboxa{{\captionfonts #1: #2}}%
  \ifdim \wd\@tempboxa >\hsize
    {\captionfonts #1: #2\par}
  \else
    \hbox to\hsize{\hfil\box\@tempboxa\hfil}%
  \fi
  \vskip\belowcaptionskip}
\makeatother   

\newcommand{\N}{\mathbf{N}}
\newcommand{\vecx}{\mathbf{x}}
\newcommand{\vecN}{\mathbf{N}}
\newcommand{\vecg}{\mathbf{g}}
\newcommand{\vecf}{\mathbf{f}}
\newcommand{\vecw}{\mathbf{w}}

\newcommand{\E}{\mathbb{E}}
\newcommand{\expect}[1]{\langle #1 \rangle}
\newcommand{\ev}[1]{\langle #1 \rangle}
\newcommand{\FPO}[1]{\mathcal{L}[#1]}
\newcommand{\R}{\mathbb{R}}
\newcommand{\cov}[0]{\text{cov}}
\newcommand{\D}{\mathcal{D}}
\newcommand{\vecv}[0]{\mathbf{v}}
\newcommand{\vecn}[0]{\mathbf{n}}
\newcommand{\vecmu}[0]{\boldsymbol{\mu}}
\newcommand{\Peff}{P_\text{eff}}
\newcommand{\vecalph}[0]{\boldsymbol{\alpha}}

\begin{document}
\hspace{13.9cm}

\ \vspace{20mm}\\

{\LARGE Particle-filtering approaches for nonlinear Bayesian decoding of neuronal spike trains}

\ \\
{\bf \large Anna Kutschireiter $^{\displaystyle 1, \displaystyle 2, \displaystyle 3}$}, {\bf \large Jean-Pascal Pfister $^{\displaystyle 1, \displaystyle 2, \displaystyle 3}$} \\
{$^{\displaystyle 1}$Institute of Neuroinformatics, UZH / ETH Zurich, Winterthurerstrasse 190, 8057 Zurich, Switzerland.}\\
{$^{\displaystyle 2}$Neuroscience Center Zurich (ZNZ), UZH / ETH Zurich, Winterthurerstrasse 190, 8057 Zurich, Switzerland.}\\
{$^{\displaystyle 3}$Department of Physiology, University of Bern, B\"{u}hlplatz 5, 3012 Bern, Switzerland.}\\
%

{\bf Keywords:} 

\thispagestyle{empty}
\markboth{}{NC instructions}
\ \vspace{-0mm}\\
%
\begin{center} {\bf Abstract} \end{center}

The number of neurons that can be simultaneously recorded doubles every seven years \cite{Stevenson2011}.
This ever increasing number of recorded neurons opens up the possibility to address new questions and extract higher dimensional signals or stimuli from the recordings.
Modeling neural spike trains as point processes, this task of extracting dynamical analog signals from spike trains is commonly set in the context of nonlinear filtering theory.
Particle filter methods relying on importance weights are generic algorithms that solve the filtering task numerically, but exhibit a serious drawback when the problem dimensionality is high:
they are known to suffer from the `curse of dimensionality' (COD), i.e. the number of particles required for a certain performance scales exponentially with the observable dimensions.
Here, we first briefly review the theory on filtering and system identification with point process observations in continuous time.
Based on this theory, we investigate both analytically and numerically the reason for the COD of weighted particle filtering approaches:
Similarly to particle filtering with continuous-time observations, the COD with point-process observations is due to the decay of effective number of particles, an effect that is stronger when the number of observable dimensions increases.
Given the success of unweighted particle filtering approaches in overcoming the COD for continuous-time observations, we introduce an unweighted particle filter for point-process observations, the spike-based Neural Particle Filter (sNPF), and show that it exhibits a similar favorable scaling as the number of dimensions grows.
Further, we derive rules for the parameters of the sNPF from a maximum likelihood approach learning, that allows both online and offline unsupervised learning.
We finally employ a simple decoding task to illustrate the capabilities of the sNPF and to highlight one possible future application of our inference and learning algorithm.


\section{Introduction}

Our nervous system represents information about the environment through coordinated spiking activity of neuronal populations.
In the last decades, multiple electrodes have become standard tools in neuroscience research, registering the spiking activity of many neurons within and across different brain regions simultaneously.
Presently, state-of-the-art probes offer the possibility to record up to 700 neurons at the same time \cite{Jun2017}.
On the one hand, this opens up new possibilities for applications, for instance in the design and implementation of brain-machine interfaces and neuroprostethic devices \cite{Donoghue2002,Shanechi2016}.
On the other hand, analysis of this immense amount of data require algorithms that scale according to this high-dimensional input.

The theory of point processes allows to model and analyse time series of discrete, possibly probabilistic events \cite{bremaud1981point}, such as the spiking behavior of neurons in response to a stimulus or a combination of stimuli.
The task of decoding is to reconstruct the stimulus from neuronal responses, which may be seen as restricted and noisy sensors.
This task naturally has to be performed by neurons, which need to estimate their environmental state, as well as by experimenters if the ground truth of the stimulus is inaccessible.
This is the case for example when decoding motor commands for prosthetics \cite{Donoghue2002} or when decoding the 'position' of an animal from place cell activities during memory recall \cite{Davidson2009}. 
Algorithms based on Bayesian statistics offer a general approach to decoding stimuli, or `states', from neuronal activity, thereby taking into account prior knowledge about the stimulus in terms of a statistical representation \cite{Brown1998,Zhang1998,Barbieri2004,Truccolo2005,Koyama2010,Macke2011,Pillow2011}.

If the stimulus is dynamic, then Bayesian decoding of the stimulus from the history of spikes can be framed in the context of filtering theory.
The formal solution to the filtering problem with point process observations is infinite dimensional and thus, beyond specific settings, impossible to write down in closed form \cite{Snyder1972,bremaud1981point}.
Algorithms relying on a Gaussian approximation of the filtering distribution are abundantly used e.g.~for decoding the position of animals from place cell activity \cite{Brown1998,Eden2007,Harel2015} (here commonly referred to: Gaussian assumed density filters, ADF).
However, these models are not able to capture multi-modal posteriors, which may occur for certain nonlinear stimulus dynamics and/or boundary conditions \cite{Huang2009}.
If the state space is discrete, a closed-form solution exists, which was used to model how filtering could be performed by a neuronal population \cite{Bobrowski2009,Legenstein2014}, and which could in principle also account for the decoding task.
However, for continuous stimuli these approaches would depend on a discretization of state-space, which is disadvantageous if it is unconstrained or if its dimensionality is large.

Particle filters (PFs, cf.~\cite{Doucet2000}) are a large class of algorithms that solve the filtering problem numerically.
By approximating the posterior with a finite number of weighted empirical samples (so-called `particles'), PFs can accommodate a wide range of state-space models, and are, unlike for instance a Kalman filter \cite{Kalman1960,Kalman1961}, not restricted to linear models.
Particle filters have received some attention in the context of spike-train decoding \cite{Brockwell2004,Ahmadian2011}, but are known to suffer from the so-called `curse of dimensionality' (COD): as the dimensionality of the problem increases, so does the demand in number of particles in order to keep the numerical performance on a satisfactory level.
For continuous-time observations, it can be shown \cite{Surace2017} that the COD manifests itself in a decreased decay time-scale of the particles' importance weights.
For point-process observations, the situation can be considered more severe, because the posterior distribution exhibits jumps.

Closely linked to, and in fact a prerequisite for effective decoding, is the encoding task, i.e.~finding the relationship of stimulus and spike generation, for instance in terms of an instantaneous firing rate.
In the context of point processes, encoding would refer to finding a model that relates the hidden state as well as the history of events with the point process observation.
Encoding has been studied extensively (for review, see \cite{Macke2011}), and mostly focuses on offline estimation \cite{Macke2011} and/or settings where the stimulus is known \cite{Eden2004}.
However, it is known that we can expect plasticity of neural dynamics even on the time scales of a single recording session \cite{Fairhall2001}.
Place cells, for instance, change their respective place fields rapidly when an animal is placed in a new environment, more slowly in response to exploration \cite{Frank2006} or as a result of external manipulation \cite{Schoenenberger2016}.
However, access to the ground-truth stimulus is not always possible, which justifies the need to formulate an encoding algorithm that is both online and unsupervised.

In this paper, we introduce an ansatz for a Bayesian decoding algorithm, the spike-based Neural Particle Filter (sNPF), that consists of a particle filter with equally weighted samples and builds on earlier work \cite{Kutschireiter2017} to account for point process observations.
The advantages of this filter are that it is formulated in continuous time and continuous-state space.
Further, in its most general form it can accommodate any nonlinear hidden dynamics and encoding model, without the need to rely on Gaussian approximations, giving it a large flexibility with respect to models and boundary conditions.
We demonstrate the performance of this filter in first a simple, one-dimensional toy model, and second, a more sophisticate model motivated by decoding the position of a foraging animal from hippocampal place cell responses.
For the latter, we show that it outperforms state-of-the-art decoding decoding methods \cite{Harel2015,Harel2016}, which are based on a Gaussian approximation, and that its performance is close to optimal.
Further, we show that in contrast to standard weighted particle filters, the sNPF is able to avoid the COD, making it a suitable candidate for decoding of high-dimensional models.
Based on a maximum-likelihood approach, the filter can be used to approximate the incomplete-data likelihood function.
From this likelihood function,  online learning rules for system identification can be derived, which may be useful for adaptive encoding when the stimulus is unknown.

\section{Problem formulation}


\subsection{Generative model}

Consider $ m $ neurons whose spiking behavior is modeled by a Poisson point process \cite{bremaud1981point} $ \vecN_t = \{N_t^{(1)},\dots,N_t^{(m)}\} $, where $ N_t^{(i)} $ counts the number of spikes of neuron $ i $ that occurred in the time interval  $ [0,t) $.
Equivalently, the trajectory of this process can be characterized by the sequence of its stochastic increments with $ dN_t^{(i)} = 1 $ if neuron $ i $ has elicited a spike at time $ t $, and $ dN_t^{(i)} = 0 $ otherwise.
These increments are drawn from a Poisson distribution:
\begin{eqnarray}
	d\mathbf{N}_t  &\sim&  \text{Poisson}(d\mathbf{N}_t; \, \mathbf{g}(\mathbf{x}_t,\N_{0:t},t)\,dt). \label{eq:Paper3 - Problem dN}
\end{eqnarray}
Here, the instantaneous firing rate $ \mathbf{g}(\mathbf{x}_t,\vecN_{0:t},t) $ captures the biophysics that enter the firing of the neuron, for instance the dependence on an external stimulus $ \vecx_t $, self-interaction and spiking history dependence such as refractoriness or interaction within the neuronal population.
Since the firing rate  $ \mathbf{g}(\mathbf{x}_t,\vecN_t,t) $ defines the relationship between the stimulus $ \vecx_t $ and the firing statistics of the neurons, it defines an encoding model, which we consider known.

We further consider a continuous-valued external stimulus $ \vecx_t \in \mathbb{R}^n $ that evolves in time, the dynamics of which obeying the It\^{o} stochastic differential equation (SDE):
\begin{eqnarray}
	d\mathbf{x}_{t}  &=& \mathbf{f} (\mathbf{x}_{t}) \, dt + \Sigma_x^{1/2}(\mathbf{x}_{t}) d\mathbf{w}_t, \label{eq:Paper3 - Problem dx}
\end{eqnarray}
where $ \vecf(\vecx_t) : \R^n \to \R^n $ denotes a potentially nonlinear drift term, $ \vecw_t \in \mathbb{R}^n $ denotes a Brownian motion process and $ \Sigma_x^{1/2} (\mathbf{x}_{t}) : \R^n \to  \mathbb{R}^{n\times n} $ is the diffusion term.
In accordance with the mathematical filtering literature, we will refer to this process as the ``(hidden) state".

As an example consider a mouse moving freely in a specified environment, while at the same time the spike train of $ m $ place cells is recorded.
The position of the mouse at time $ t $ is given by $ \vecx_t $, which gives rise to the place cell activity via the instantaneous firing rate $ \mathbf{g}(\mathbf{x}_t,\vecN_{0:t},t) $.
In the simplest case of no neuronal interaction, $ g^{(i)}(\mathbf{x}_t,\vecN_{0:t},t) = \lambda^{(i)}(\vecx_t)$ would correspond to the place field or spatial tuning curve of neuron $ i $.
The spike train corresponds to the \emph{observations}.
The dynamics of the mouse is given in Eq.~\eqref{eq:Paper3 - Problem dx}.
Here, the stochastic diffusion term models the erratic behavior of the mouse, e.g. during forage, and the nonlinearity in the drift term $ \vecf(\vecx_t) $ may model specific attractors or repellers in the environment, such as food sources or averse stimuli.
If the drift term is zero, the mouse performs a random walk.
Note that for certain drift functions $ \vecf(\vecx_t) $\footnote{E.g.~with a negative leading order in the components.} or spatial boundary conditions, Eq.~\eqref{eq:Paper3 - Problem dx} gives rise to a stationary probability distribution, which, together with the time scale of the movement, serves as \emph{prior} knowledge.
We will elaborate on this toy model further in section \ref{sec:Decoding position from place cell activity}. 

\subsection{Decoding}

For this example, the (online) decoding task would read:
given the knowledge about the dynamics of the mouse, and given the knowledge about the place fields, what is the position of the mouse at time $ t $ given the spike trains of the neurons up to this time.

More formally, we can frame decoding a nonlinear filtering problem:
Given the model in Eqs.~\eqref{eq:Paper3 - Problem dN} and \eqref{eq:Paper3 - Problem dx} and given the sequence of point-process observations $ \N_{0:t} $, what is the \emph{posterior} probability distribution (or filtering distribution) $ p(\vecx_t|\N_{0:t}) $.
Equivalently, a full probabilistic solution to the filtering problem is known if the posterior expectation of any real-valued, measurable function $ \phi $ can be computed:
\begin{eqnarray}
	\expect{\phi(\vecx_t) } :=	\mathbb{E}[\phi(\vecx_t) | \N_{0:t} ]  & = & \int d\vecx_t \, \phi(\vecx_t) p(\vecx_t|\N_{0:t}) .
\end{eqnarray}
A formal solution for the evolution of this posterior expectation $ \expect{ \phi_t }  $\footnote{For brevity, here we use superscripts to denote vector components and we write $ \phi(\vecx_t) = \phi_t $ and $ g_d (\vecx_t) = g_{d,t} $ whenever there is no ambiguity.} can be written down in terms of an SDE \cite[T7, p.~173]{bremaud1981point}:
\begin{eqnarray}
	d\expect{ \phi_t } & = & \expect{\FPO{\phi_t}} \, dt +   \cov(\phi_t,\vecg_t^T) \text{diag} \left( \expect{\vecg_t} \right) ^{-1} \left( d\N_t - \expect{\vecg_t} \, dt \right) , \label{eq:Paper 3 - posterior expect formal}
\end{eqnarray}
where $  \FPO{\phi_t} := \sum_i f_i(\vecx_t)  \frac{\partial }{\partial x_i} \phi_t + \sum_{i,j} (\Sigma_x^{1/2})_{ij} \frac{\partial^2}{\partial x_i \partial x_j } \phi_t $ is the infinitesimal generator of the hidden process in Eq.~\eqref{eq:Paper3 - Problem dx}.
This solution suffers from a closure problem, i.e.~it is analytically intractable, which can be seen when writing down the SDE for the first posterior moment, with $ \phi(\vecx_t) = \vecx_t $:
\begin{eqnarray}
	d\expect{\vecx_t} & = &\expect{\vecf(\vecx_t)} \, dt +  \cov( \vecx_t,\vecg_t^T) \text{diag} \left( \expect{\vecg_t} \right) ^{-1} \left( d\N_t - \expect{\vecg_t} \, dt \right), \label{eq:Problem first moment}
\end{eqnarray}
which clearly depends on higher-order posterior moments.
In general, any SDE for posterior moments will depend on higher-order moments.

Note that filtering usually asks for the full statistical description, while for decoding one would have to specify how the state estimate $ \hat{\vecx}_t $ is obtained from this description\footnote{For example, Pillow et al.~\cite{Pillow2011} consider the maximum \emph{a posteriori}, i.e.~the mode of the filtering distribution, whereas Harel et al.~\cite{Harel2015} consider the first moment. For these models it does not make a difference, because they both approximate the filtering distribution by a Gaussian, where mode and first moment coincide. In a more general model, for instance with a multimodal posterior, one would need to specify which of these is considered to be the decoding estimate $ \hat{\vecx}_t $, because in general these are not the same.}.
Here, by decoding we refer to determining the full posterior density of the state $ \vecx_t $.

\section{The sNPF: A particle-based decoding algorithm}

We propose that the decoding problem defined by Eqs.~\eqref{eq:Paper3 - Problem dx} and \eqref{eq:Paper3 - Problem dN} can approximately be solved by considering $ P $ realizations from the It\^{o} SDE:
\begin{eqnarray}
	d \mathbf{x}_t^{(i)} &=& \mathbf{f}(\mathbf{x}_t^{(i)})\,dt + \Sigma_x^{1/2} \, d\vecw^{(i)}_t +  W_t \left( d\mathbf{N}_t - \mathbf{g}(\mathbf{x}_t^{(i)},\vecN_{0:t},t)\,dt \right),  \label{eq:Paper 3 - sNPF}
\end{eqnarray}
where $ i=1\dots P $ denotes the realization, $ \vecw_t^{(i)} $ denote independent vector Brownian motion processes and $ W_t $ is a time-dependent gain matrix.
we call this ansatz for the particle dynamics the Neural Particle Filter(sNPF), following a similar ansatz for continuous-time observations (Neural Particle Filter, \cite{Kutschireiter2017}).

Equation \eqref{eq:Paper 3 - sNPF} suggests an approximation of the posterior density via
\begin{eqnarray}
	p(\vecx_t|\N_t) & \approx & \frac{1}{P} \sum_{i=1}^{P} \delta(\vecx_t - \vecx_t^{(i)} ), \label{eq:Paper 3 - sNPF - empirical density}
\end{eqnarray}
where posterior expectations of a scalar-valued function $ \phi(\vecx_t)  $ are approximated with
\begin{eqnarray}
	\mathbb{E}[\phi(\vecx_t) | \N_t ] \approx \expect{\phi_t} := \frac{1}{P} \sum_{i=1}^P \phi (\vecx_t^{(i)}). \label{eq:Paper 3 - sNPF particle representation}
\end{eqnarray}
In this way, Eq.~\eqref{eq:Paper 3 - sNPF} defines a particle-filtering algorithm, where each of the realizations of the SDE corresponds to a single particle trajectory.
Due to the particle representation, the sNPF is able to account for multimodal posterior densities, unlike standard approaches in Bayesian decoding, such as the Gaussian assumed density filter (ADF, cf.~\cite{Brown1998,Harel2015}).

Our ansatz is motivated by the formal solution to the nonlinear filtering problem with point-process observations (\cite{bremaud1981point}, for details see Section S1 in the SI, cf.~also Eq.~\ref{eq:Paper 3 - posterior expect formal}).
The derivation of the formal solution remains unaffected by a possible history dependence of the firing rate function, and thus the same holds for our ansatz.
It is governed by both the stimulus dynamics serving as a prior ('prediction'), as well as the difference between actual point process observations and a single-particle rate estimate ('correction'), modulated by a time-dependent gain $ W_t $.
Similarly to a Kalman gain \cite{Kalman1960,Kalman1961}, it determines the impact of the point-process observations on the particle trajectories. 
If it is large, the prior dynamics will be negligible and the posterior will be determined mostly by the observations.
On the other hand, if it is small, the observations will hardly affect the particle trajectories.
In this case, the posterior will be given by the stationary solution of Eq.~\eqref{eq:Paper3 - Problem dx}.


\subsection{Determining the gain matrix}

How well the particle representation matches the true posterior distribution is mainly determined by two factors: first, the number of particles, and second, the gain matrix $ W_t $ and how it is determined.
Here, we present two possibilities to determine the gain.

The first one, like the ansatz in Eq.~\eqref{eq:Paper 3 - sNPF}, is motivated by the formal solution to the filtering problem (in particular the SDE of the first moment in Eq.~\ref{eq:Problem first moment}).
Here, the gain matrix $ W_t$ is computed according to $ W_t = \cov \left( \vecx_{t}, \vecg_t^T \right) \text{diag} \left( \expect{\vecg_t} \right)^{-1}  $, where the covariance between state $ \vecx_t $ and instantaneous firing rate $ \vecg_t $ as well as posterior expectations are empirically estimated from the set of particles.
In this case, the resulting filter is termed ``Ensemble Kushner-Stratonovich-Poisson Filter" (EKSPF, \cite{Venugopal2015}).
Even though this filter is perfectly suited for applications like decoding neuronal spike trains in a straightforward way, it has not received any attention yet in this regard. 
In Section \ref{sec:Decoding position from place cell activity} we will use the EKSPF for decoding.

Alternatively, we may adopt a completely different viewpoint on the gain matrix $ W_t $:
instead of determining it from the particle positions at each time step, we consider it an adaptive decoding parameter, i.e.~a parameter that can be determined online with a maximum likelihood framework.
This will be outlined in Section \ref{sec:Adaptive Encoding} below.


\section{Decoding position from place cell activity - a simulation study} \label{sec:Decoding position from place cell activity}

\label{sec:Paper 3 - Decoding position from place cell activity}

In this section, we will demonstrate the capabilities of the sNPF using a toy model that is motivated by the task of decoding the position of an animal based on spike trains in place cells.
We will compare its performance with that of established filtering algorithms, in particular the Bootstrap particle filter (BPF, \cite{Doucet2000}, see Section S~4 in the SI) and the Gaussian ADF \cite{Harel2016} (see Section S~2.2 in the SI for details).
Note that the dimensionality of the problems shown here is still small enough to have almost-optimal performance of the BPF with a computationally reasonable number of particles (here, we choose $ P = 1000 $ for all simulations shown).
Hence, we can consider simulation results of the BPF as the `ground truth' benchmark.

Place cells in the hippocampus are known to exhibit position-dependent responses when the animal (for instance a mouse) is placed in a confined box, which can be modeled by a Gaussian-shaped tuning curve (cf.~\cite{Brown1998}).
Thus, the firing rate function in our model reads
\begin{eqnarray}
g_d(\vecx_{t}) & = & g_0 \exp( -\frac{1}{2} ( \vecx_{t} - \boldsymbol{\mu}_d )^T \Sigma_d^{-1}  ( \vecx_{t} - \boldsymbol{\mu}_d ) ), \label{eq:Paper 3 - g_d(x)}
\end{eqnarray}
where $ g_0 $ denotes the maximal firing rate, and $ \boldsymbol{\mu}_d $ and $ \Sigma_d $ denote the tuning-curve center and variance of place cell $ d $.


\subsection{One-dimensional toy model}

\label{sec:Paper 3 - One-dimensional toy model}

\begin{figure}
	\centering
	\begin{subfigure}{0.49\textwidth}
		\caption{\label{fig:Paper3 - 1D place cell decoding - placefields}}
		\includegraphics[width=\textwidth]{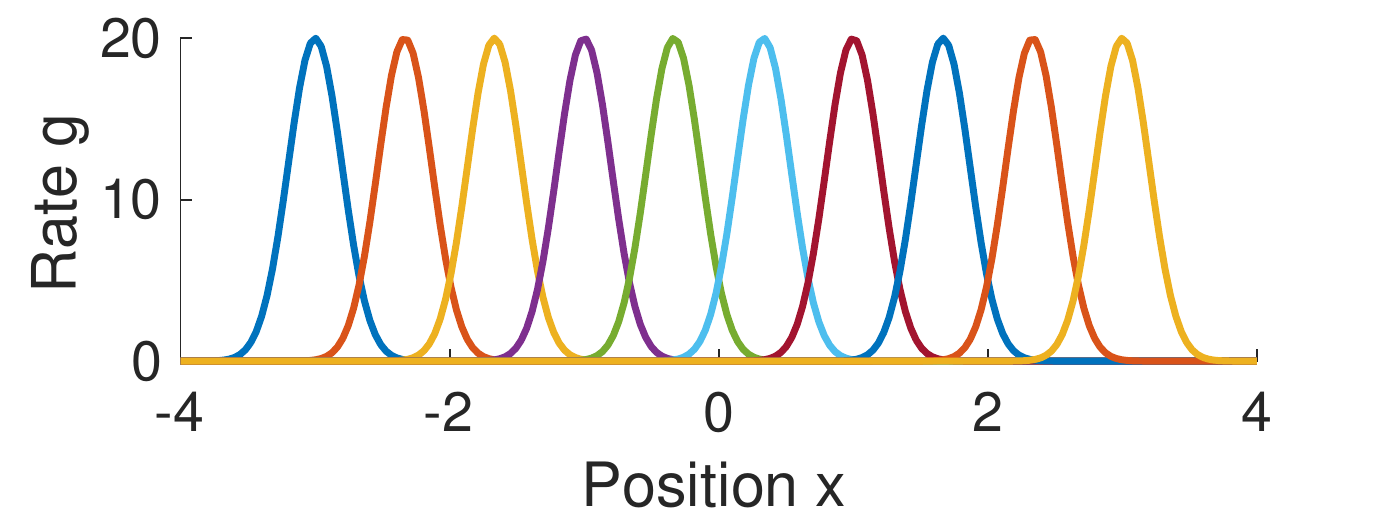}
	\end{subfigure}
	\begin{subfigure}{0.49\textwidth}
		\caption{\label{fig:Paper3 - 1D place cell decoding - sNPF vs ML}}
		\includegraphics[width=\textwidth]{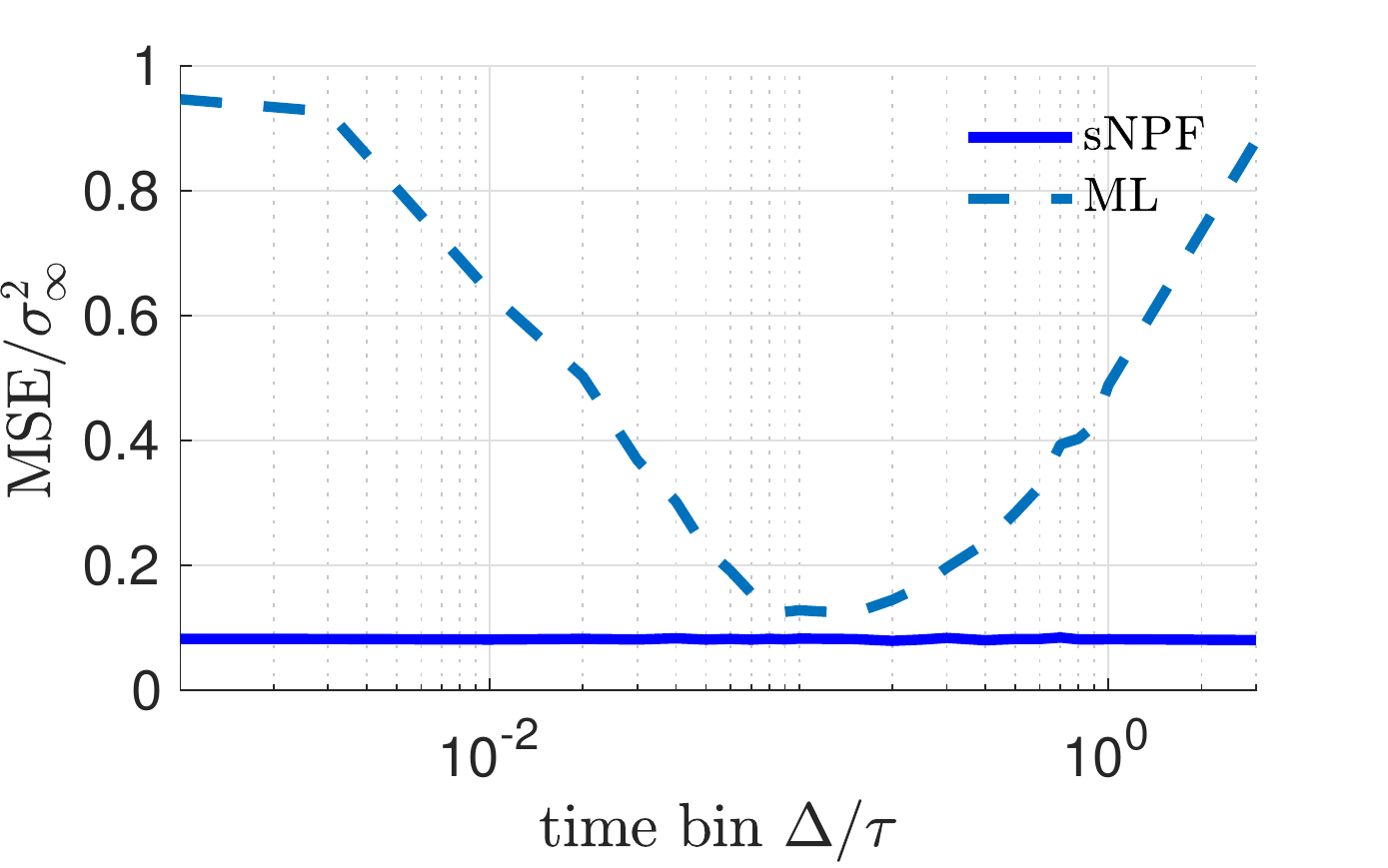}
	\end{subfigure}
	\begin{minipage}{0.49\textwidth}
		\begin{subfigure}{\textwidth}
			\caption{Linear hidden dynamics.
				\label{fig:Paper3 - 1D place cell decoding - OU}}
			\includegraphics[width=\textwidth]{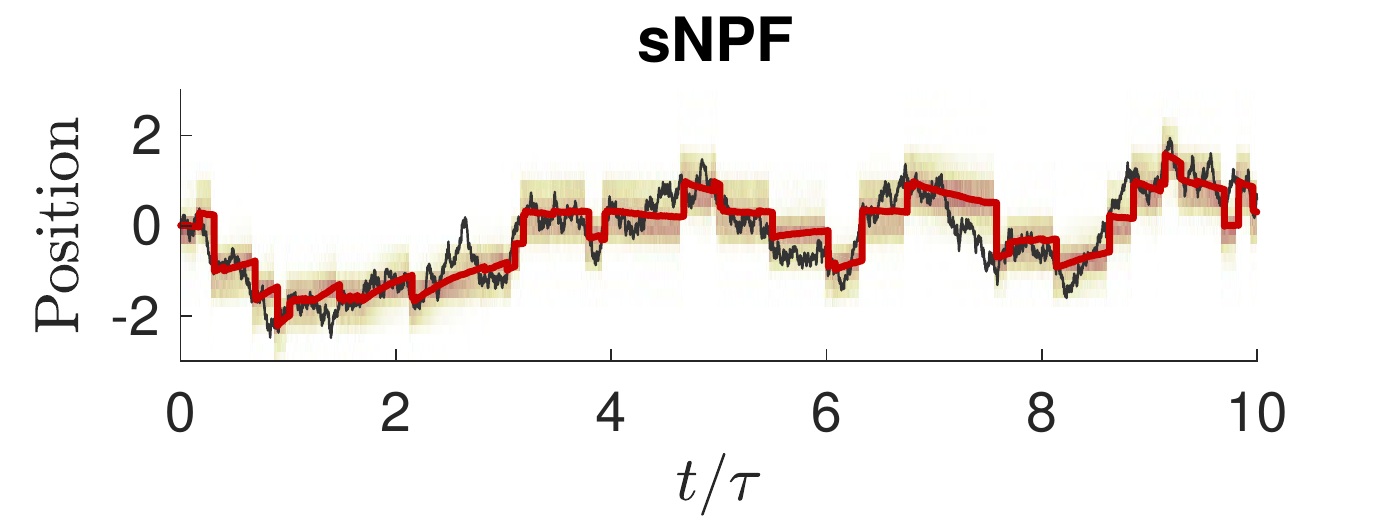}
			\includegraphics[width=\textwidth]{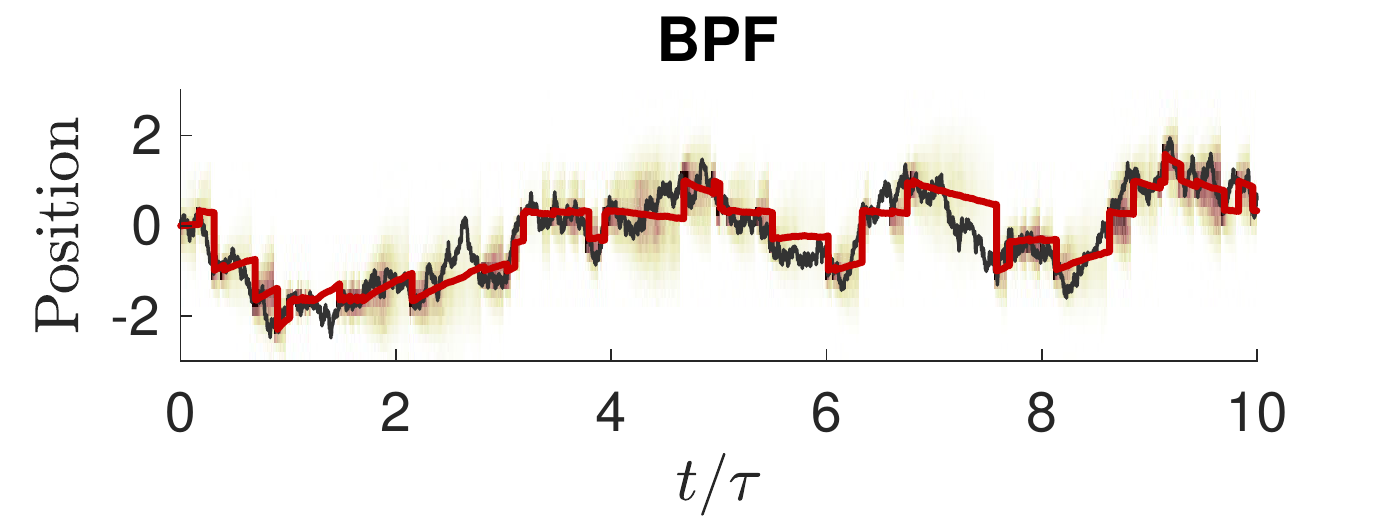}
			\includegraphics[width=\textwidth]{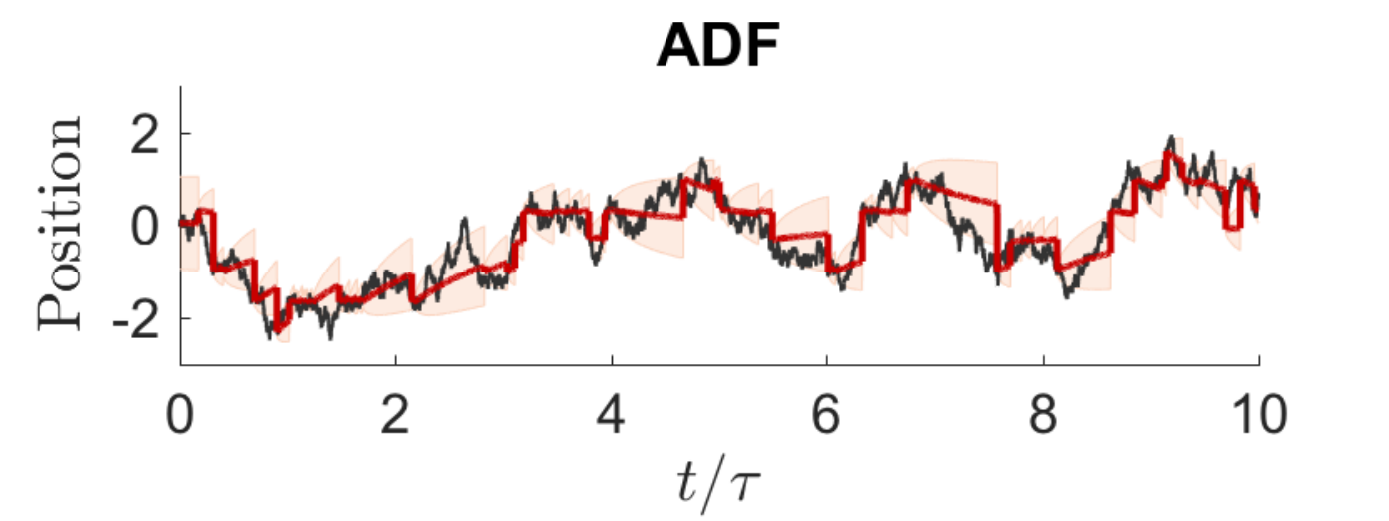}
			\includegraphics[width=\textwidth]{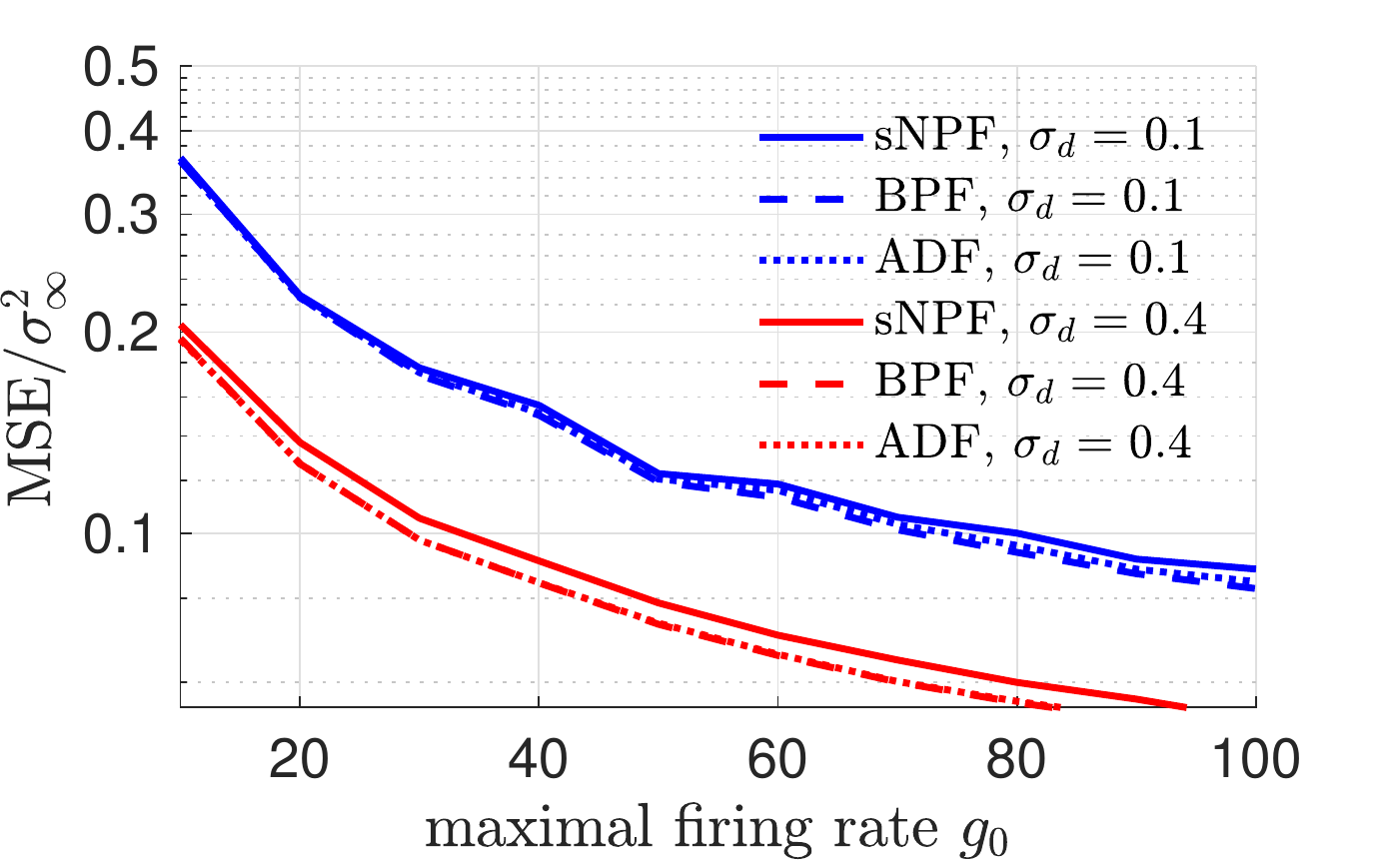}
			
		\end{subfigure}
	\end{minipage}
	\begin{subfigure}{0.49\textwidth}
		\caption{Nonlinear hidden dynamics.
			\label{fig:Paper3 - 1D place cell decoding - nonlinear}}
		\includegraphics[width=\textwidth]{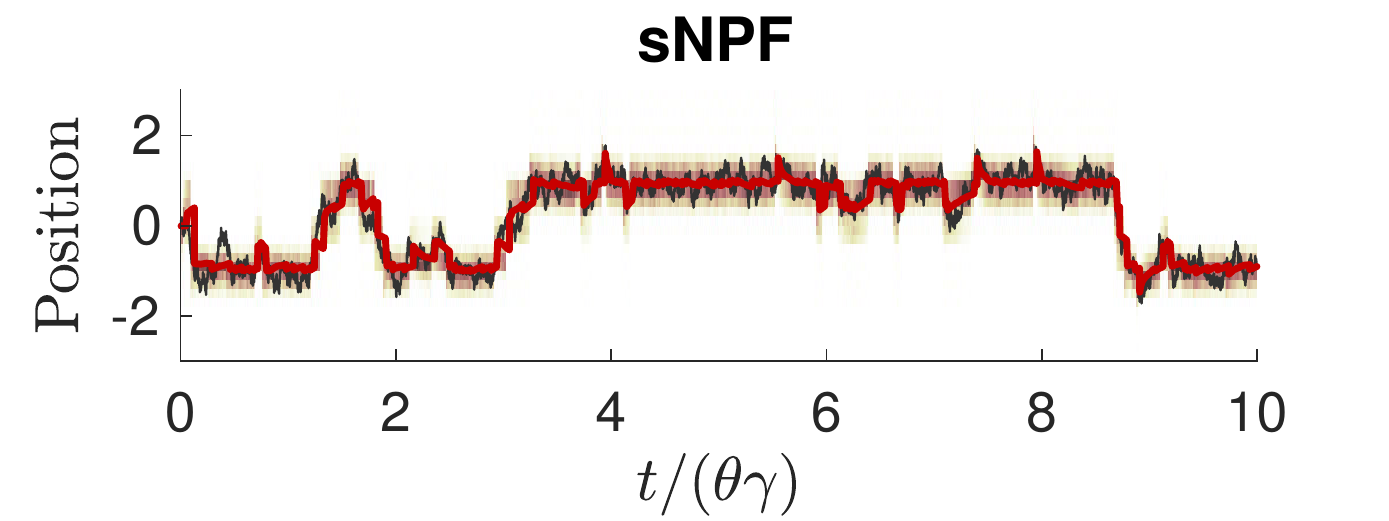}
		\includegraphics[width=\textwidth]{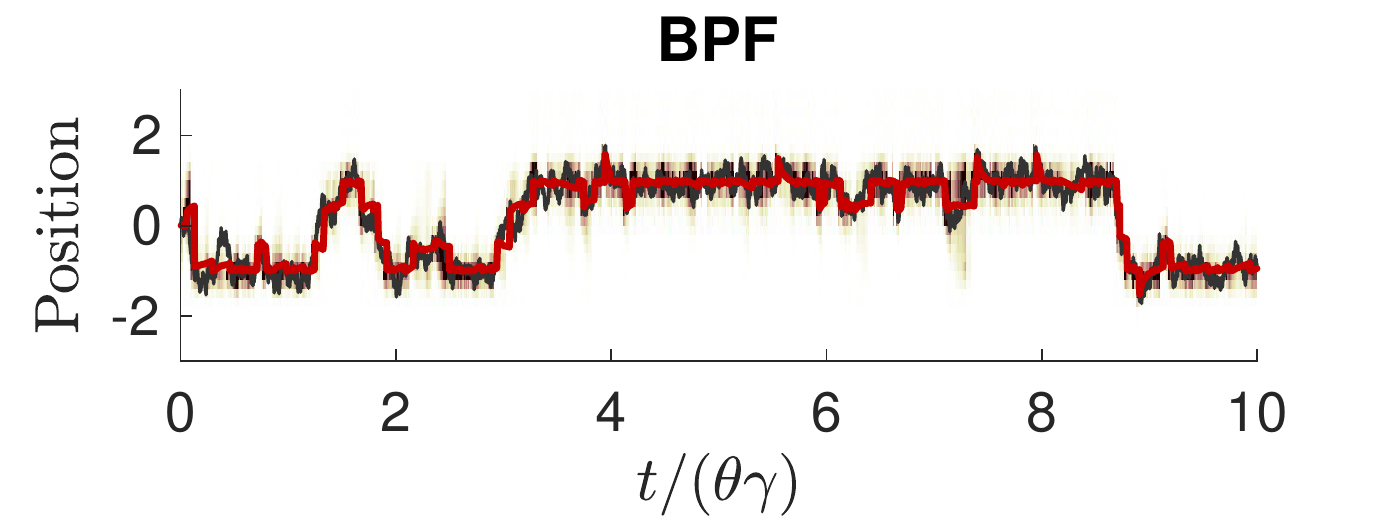}
		\includegraphics[width=\textwidth]{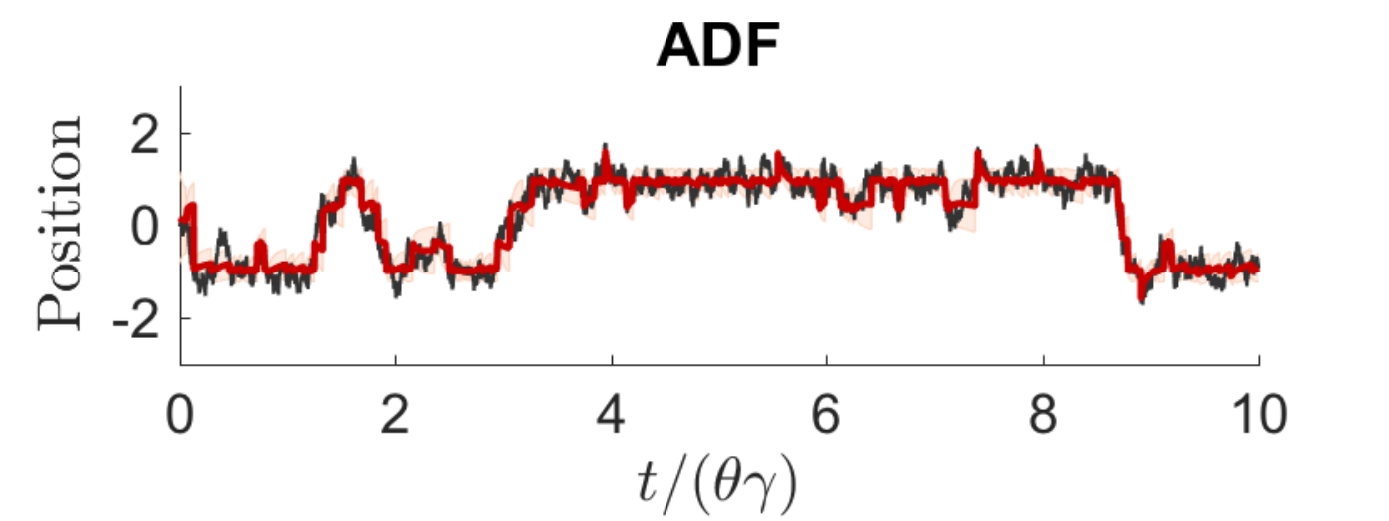}
		\includegraphics[width=\textwidth]{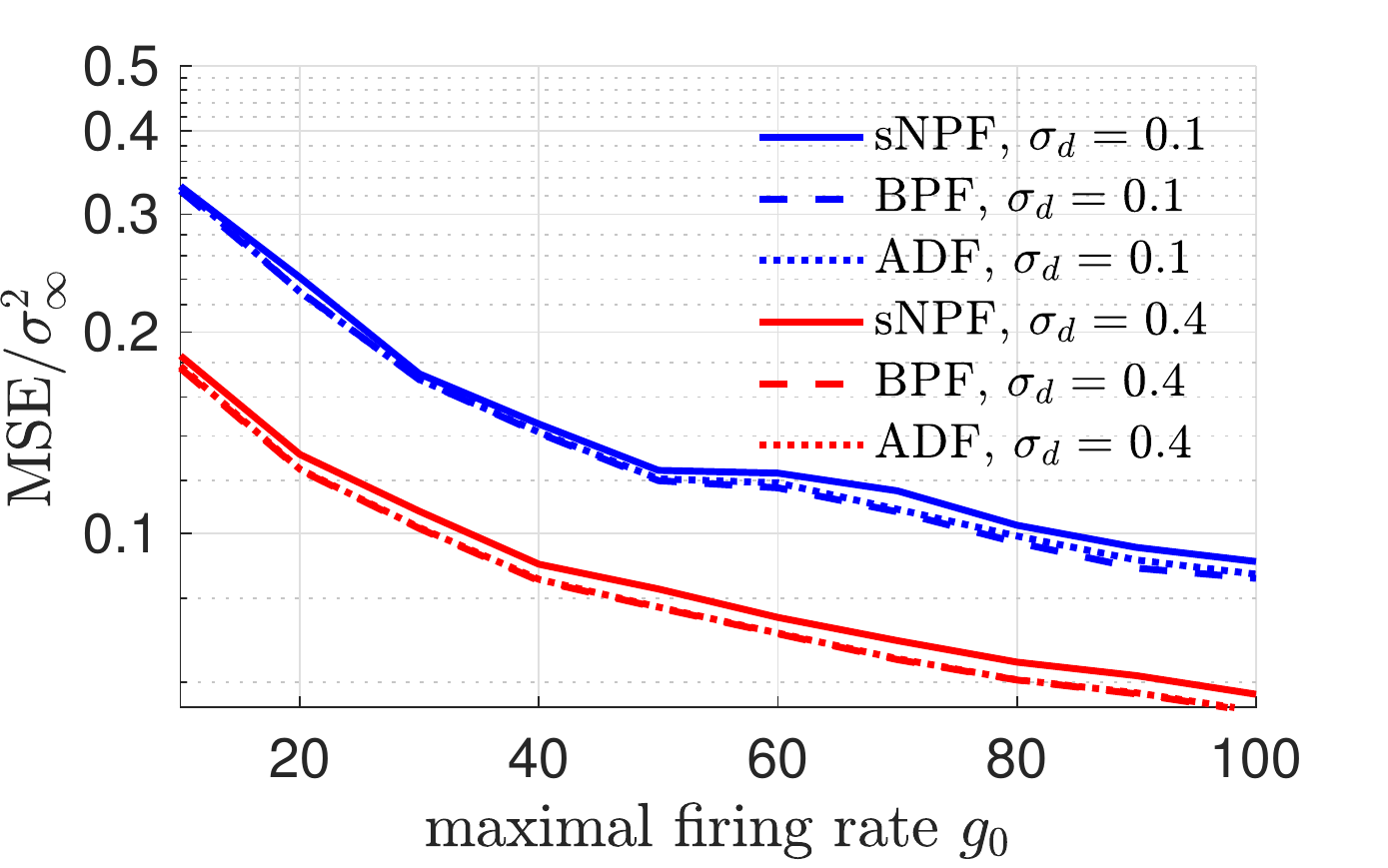}
	\end{subfigure}
	\begin{minipage}{0.49\textwidth}
		
	\end{minipage}
	\caption[One-dimensional place-cell decoding.]{
		\label{fig:Paper3 - 1D place cell decoding}
		{\bf One-dimensional place-cell decoding.}
		{\bf{(a)}} 
		Illustration of place fields. Here, $ m=10 $ neurons with place field centers $ \mu_d $ spaced  equally between -3 and 3, place field width $ \sigma_d = 0.2 $ and maximal firing rate $ g_0 = 20 $.
		{\bf{(b)}}
		Comparison of filtering performance of ML decoder versus sNPF as a function of time bin size $ \Delta $.
		The hidden dynamics follows an OU process (Eq.~\ref{eq:Paper 3 - 1D toy model - OU}) with $ \tau = 1 $ and $\sigma_x = \sqrt{2}$.
		{\bf{(c)}}
		Tracking simulation of the sNPF, BPF and ADF.
		The hidden dynamics is given by the black line (OU process) and the red line corresponds to the estimated first posterior moment of the corresponding filter.
		For the particle filters, shadings correspond to (weighted) particle density and thus estimated posterior density, with regions of high density appearing darker than regions of low density.
		For the ADF, the shaded regions denote the square root of the estimated posterior moment, i.e.~$ \sqrt{\Sigma_t} $.
		The lower panel shows the filtering performance in terms of MSE for different place field widths and maximal firing rates.
		The plots for BPF and ADF are almost indistinguishable by eye, and performance is only slightly better than the sNPF.
		{\bf{(c)}}
		Same as (b), but with a nonlinear hidden dynamics (Eq.~\eqref{eq:Paper 3 - 1D toy model - bimodal} with $ \theta = -3 $, $ \gamma = 1 $ and $ \sigma_x = \sqrt{2} $).
	}
\end{figure}

We first consider decoding from a one-dimensional toy model with $ d=10 $ model neurons.
These neurons have equal place-field width and maximal firing rate $ g_0 $, and their place-field centers are distributed equally within a certain range on the $ x $ axis.  (Figure \ref{fig:Paper3 - 1D place cell decoding - placefields}).

In this example, we consider both a linear hidden dynamics, i.e.~an Ornstein-Uhlenbeck (OU) process with time scale $ \tau $:
\begin{eqnarray}
dx_t & = & -\frac{1}{\tau} x_t \, dt + \sigma_x dw_t, \label{eq:Paper 3 - 1D toy model - OU}
\end{eqnarray}
which gives rise to a Gaussian stationary (prior) density with mean $ \mu_{x,\infty} = 0 $ and variance $ \sigma_{x,\infty} =  \frac{1}{2} \sigma_x^2 \tau $.
As an alternative, we also consider a nonlinear hidden dynamics:
\begin{eqnarray} 
dx_t & = & \theta  x_t ( \gamma -  x_t^2 ) \, dt + \sigma_x dw_t. \label{eq:Paper 3 - 1D toy model - bimodal}
\end{eqnarray}
The resulting stationary distribution is symmetric and bimodal, with two distinct modes at $ x = \pm \gamma $.

For the linear dynamics, we first compare the filtering performance of the sNPF in terms of its mean-squared error (MSE)  to that of a simple maximum likelihood (ML) decoder (see Section S~2.1 for details).
This decoder does not take into account any prior knowledge about the hidden process $ \vecx_{t} $, and thus simply relies on the information given by the spikes trains.
Further, its performance is heavily dependent on temporal binning of the spike train (Figure \ref{fig:Paper3 - 1D place cell decoding - sNPF vs ML}), which makes it difficult to use practically.
Certainly, the optimal bin size for decoding depends on the effective firing rate of the neurons as well as on the hidden dynamics, and is difficult to determine analytically\footnote{In a way, the optimal bin size does incorporate prior knowledge about the hidden dynamics indirectly.}. 
Using a filtering algorithm such as the sNPF instead has the advantage of being independent of hand-tuned temporal binning, and for the example simulation shown in Figure \ref{fig:Paper3 - 1D place cell decoding - sNPF vs ML} yields a much better performance than the ML decoder.

We further compared the sNPF to other \emph{filtering} algorithms, i.e.~the BPF and the ADF, which by definition \emph{do} take into account prior knowledge in terms of the hidden dynamics given by Eq.~\eqref{eq:Paper3 - Problem dx}.
In Figure \ref{fig:Paper3 - 1D place cell decoding}b and c, we show sample tracking simulations from these algorithms for both the linear and the nonlinear model, as well as a comparison of filtering performance.
For the linear model, the performance of the ADF is indistinguishable from that of the BPF.
This implies that the Gaussian approximation needed for the ADF to work is indeed justified for this particular model.
Interestingly, the same applies for the nonlinear model, even though one would have expected the ADF to perform worse due to the nonlinearity.
That its MSE is almost indistinguishable from that of the BPF demonstrates that for the parameters tested here, the Gaussian approximation is still valid.
However, the picture changes slightly if the effective rate of point emission events is decreased, e.g.~if there are only a few neurons with relatively low firing rate.
Then, the posterior becomes bimodal, and multimodality cannot be captured by the assumed-density filter (ADF) that by construction relies on a unimodal approximation.

Even though the sNPF is a very heuristic approximation, and particle dynamics relies on the dynamics of only the first moment, the sNPF performs almost as good as the benchmark.
What is notable, however, is that apparently higher-order moments fall short when compared to that of the benchmark filters, in particular in intermediate effective firing rate regimes (compare for instance the discussion on the second moment of the sNPF in Section S~3 of the SI).


\subsection{Restricted Brownian motion model}

\label{sec:Paper 3 - 2D RBM model}

Let us now consider a more realistic toy example of an animal moving in a two-dimensional, confined box $ \D $, with $ d = 25 $ spatially-tuned neurons.
Taking into account that animal movements are usually rather smooth, we model the movement within the box as an integrated Brownian motion, i.e.~we augment the hidden space by a velocity component $ \vecv_t $.
Thus, the hidden dynamics are given by a reflected stochastic differential equation (RSDE, cf.~\cite{Hanks2017}):
\begin{eqnarray}
d \vecx_t & = & \vecv_t \, dt + \mathbf{k}_t \, dt, \label{eq:Paper 3 - 2d model dx}
\\
d \vecv_t & = & - \beta ( \vecv_t - \boldsymbol{\psi} (\vecx_t)) \, dt + \Sigma_v^{1/2} d\vecw_{v,t}, \label{eq:Paper 3 - 2d model dv}
\end{eqnarray}
resulting in the hidden state $ [\vecx_t,\vecv_t]^T \in \R^4 $.
Here, $ \beta $ is a constant, $ \vecw_{v,t} \in \R^2 $ is a 2-dimensional Brownian-motion process and $\boldsymbol{\psi} (\vecx_t)$ can be thought of as the negative gradient of a potential surface, e.g.~$  \boldsymbol{\psi} (\vecx_t) = - \nabla_\vecx H(\vecx_t) $.
For simplicity, in our simulation we will use a linear function $ \boldsymbol{\psi} = - \theta \vecx_{t} $, but in general we would assume that animal movement is governed by a much more complicated nonlinearity \cite{Huang2009}.

\begin{figure}
	\centering
	\begin{subfigure}{0.39\textwidth}
		\caption{\label{fig:Paper3 - 2D place cell decoding - hidden trajectory}}
		\includegraphics[width=\textwidth]{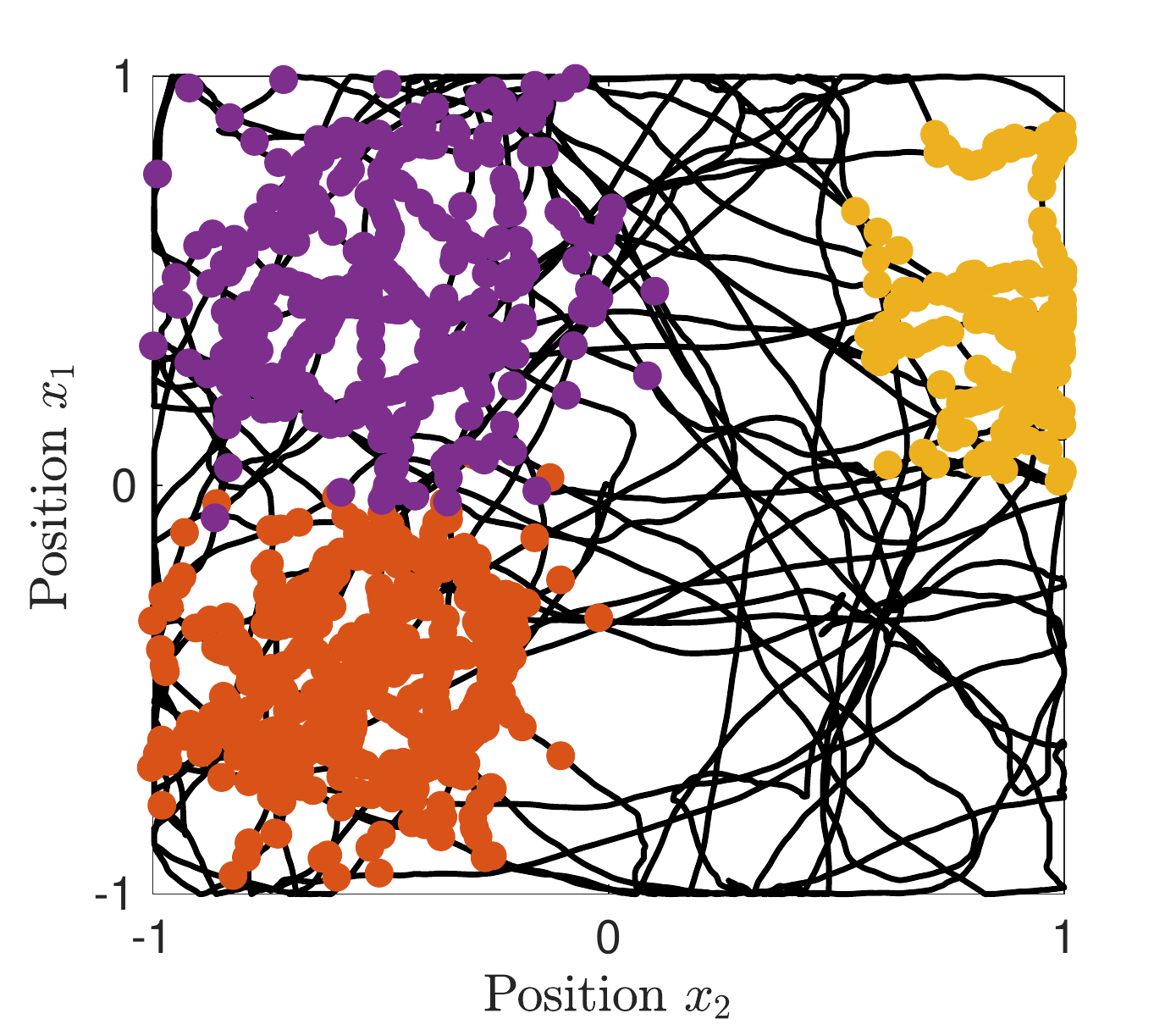}
	\end{subfigure}
	\begin{subfigure}{0.6\textwidth}
		\caption{\label{fig:Paper3 - 2D place cell decoding - filter trajectory}}
		\includegraphics[width=\textwidth]{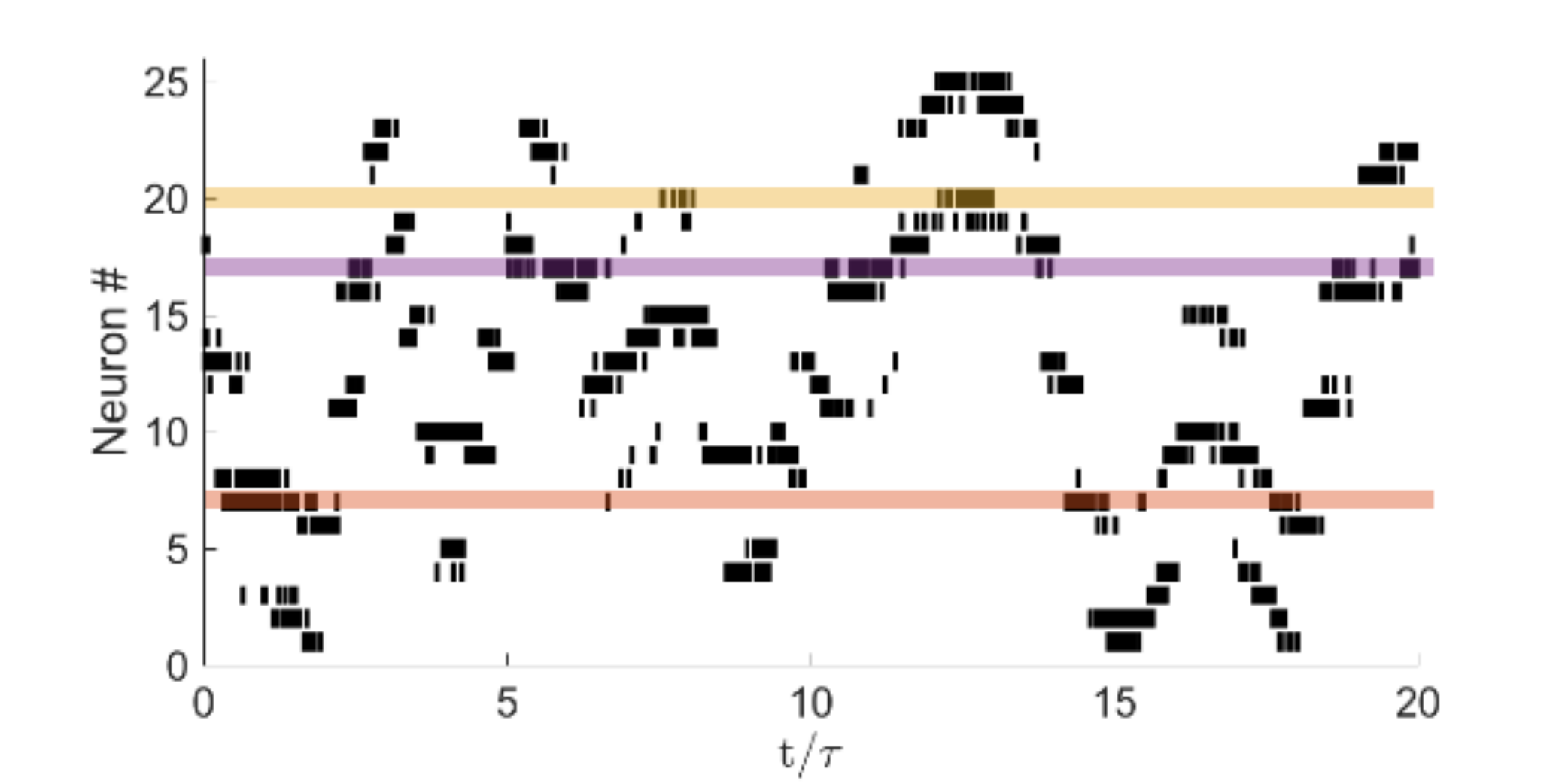}
	\end{subfigure} 
	\\
	\begin{subfigure}[p]{\textwidth}
		\caption{}
		\centering
		\includegraphics[width=0.31\textwidth]{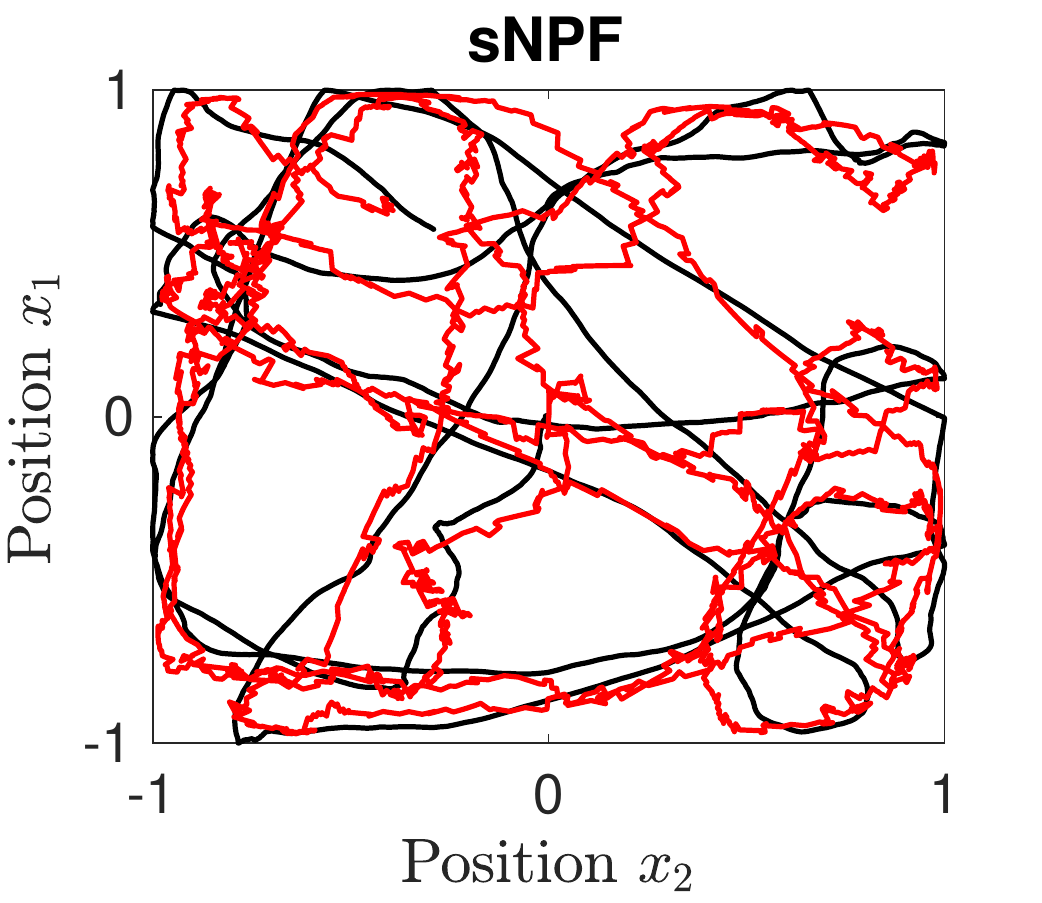}
		\includegraphics[width=0.31\textwidth]{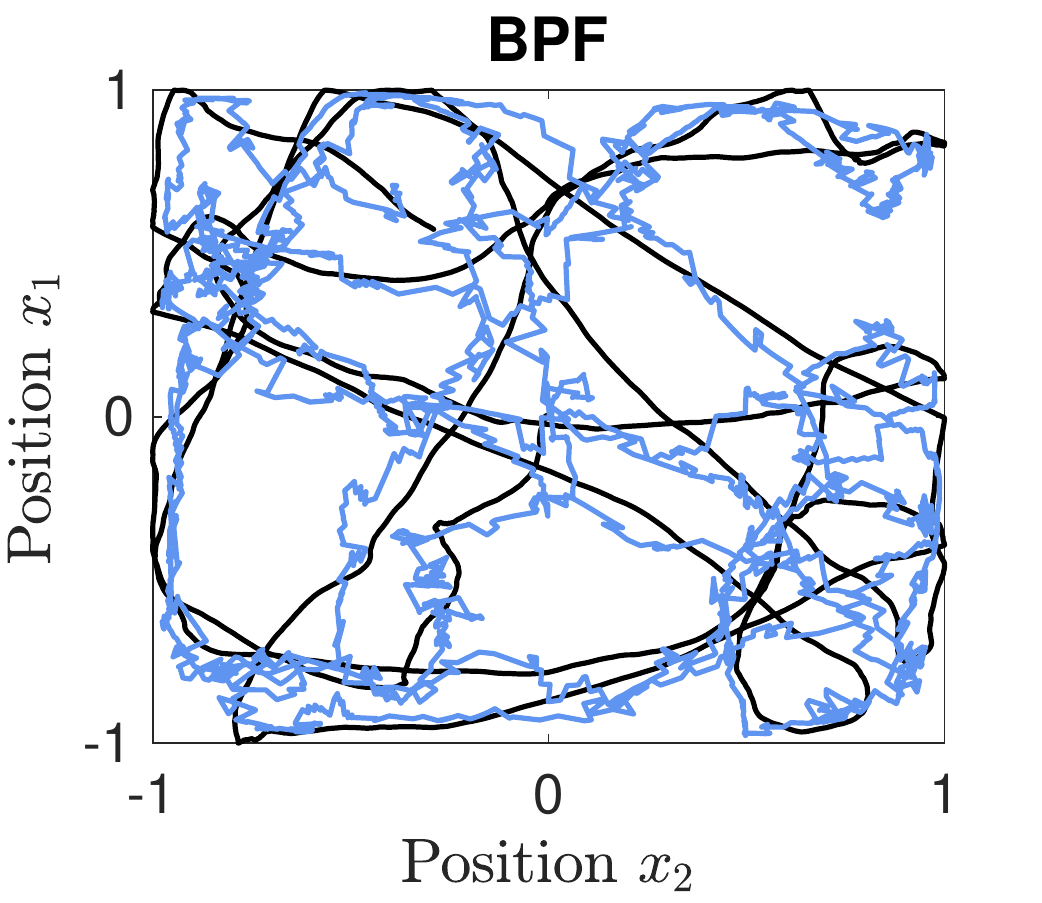}
		\includegraphics[width=0.33\textwidth]{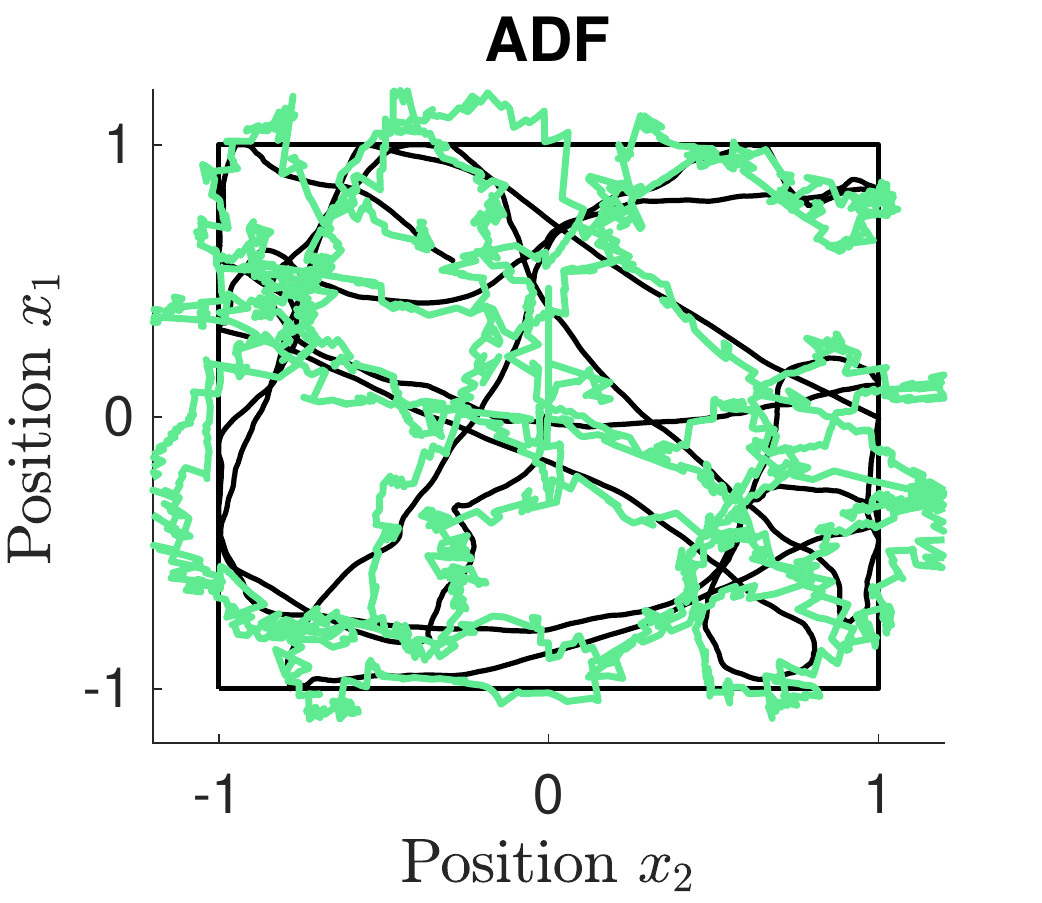}
	\end{subfigure} 
	\\
	\begin{subfigure}{0.59\textwidth}
		\caption{\label{fig:Paper3 - 2D place cell decoding - MSE}}
		\includegraphics[width=\textwidth]{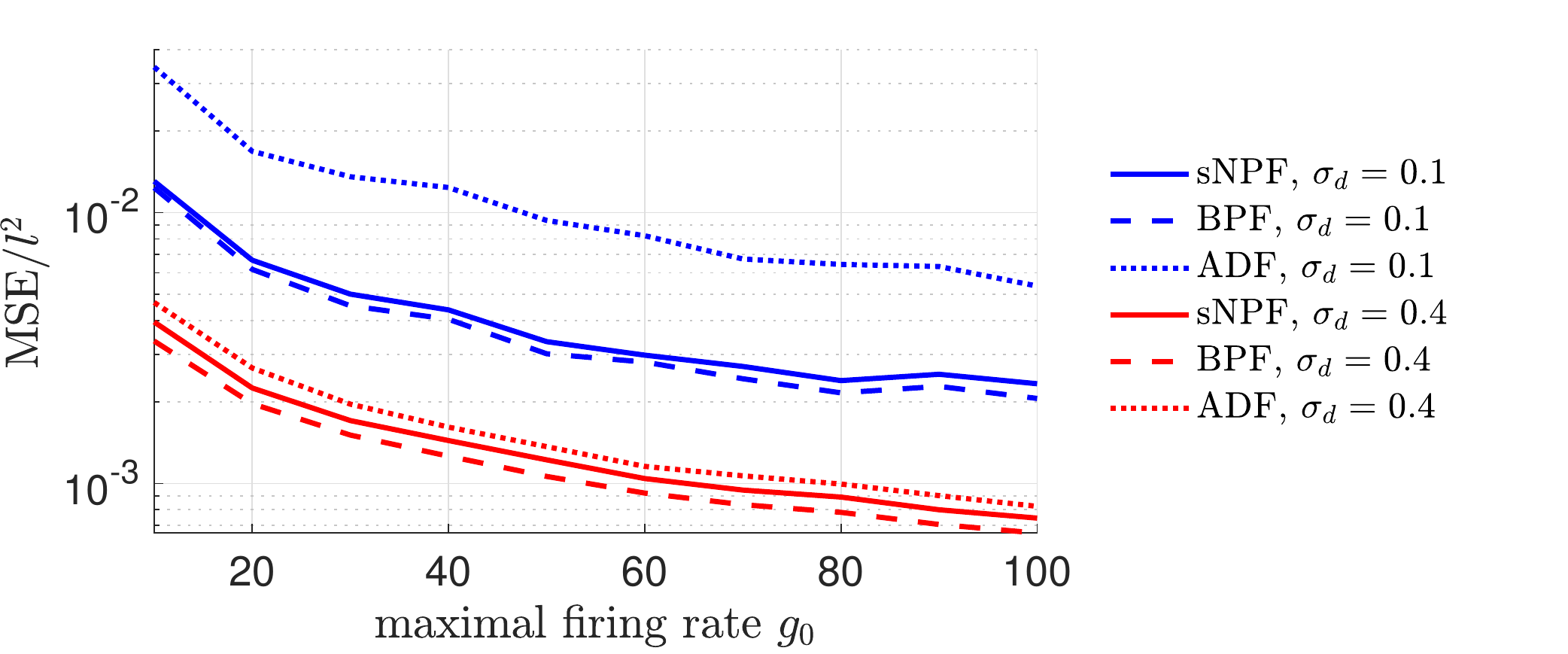}
	\end{subfigure} 
	\begin{subfigure}{0.39\textwidth}
		\caption{\label{fig:Paper3 - 2D place cell decoding - MSE harder fail}}
		\includegraphics[width=\textwidth]{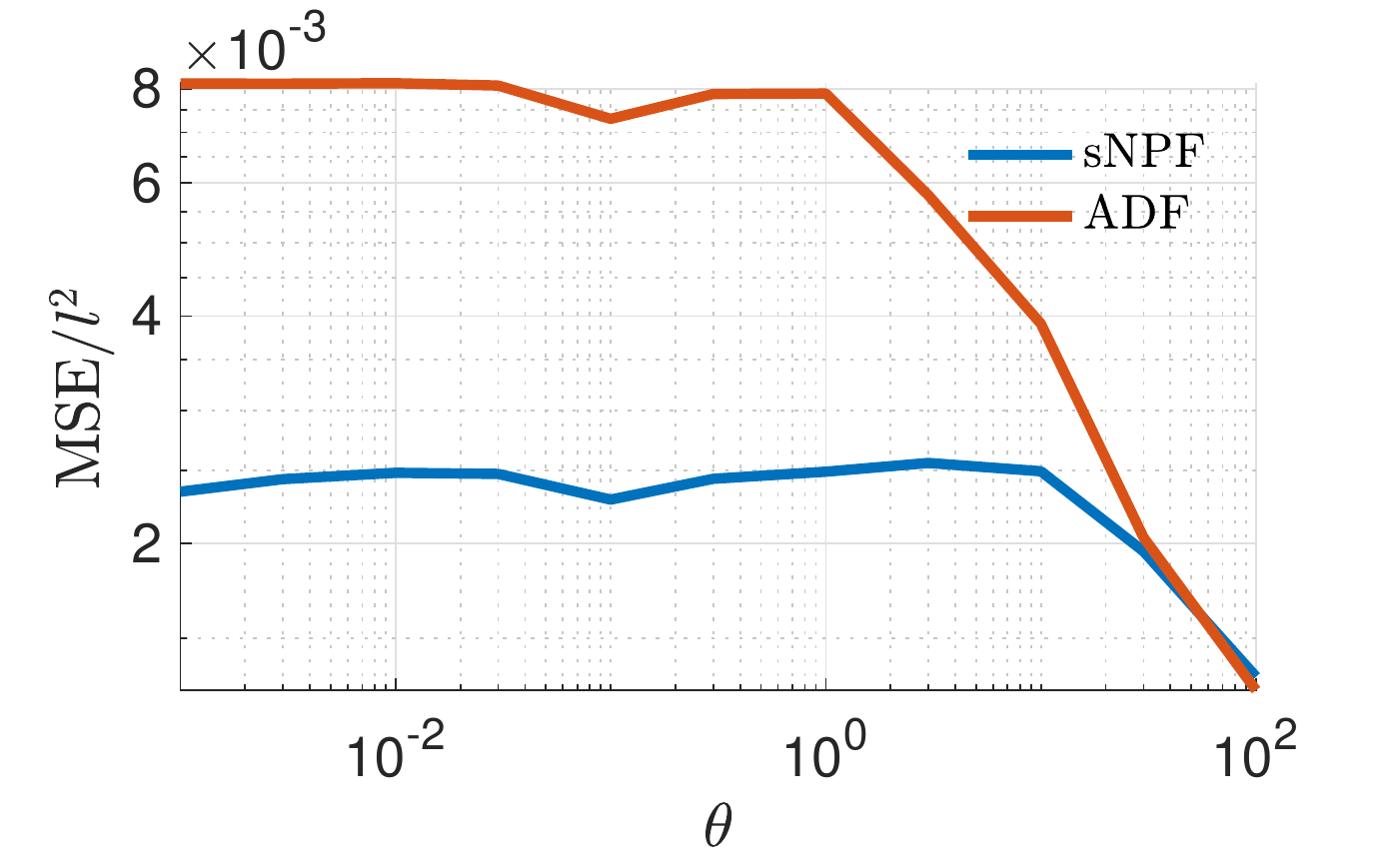}
	\end{subfigure} 
	\caption[Two-dimensional place-cell decoding in confined region.]{
		\label{fig:Paper3 - 2D place cell decoding}
		{\bf Two-dimensional place-cell decoding in confined region.}
		\textbf{(a)} Sample trajectory (spatial components) of hidden process in Eq.~\eqref{eq:Paper 3 - 2d model dx} and \eqref{eq:Paper 3 - 2d model dv}, and responses of 3 (out of $ d = 25 $) spatially tuned sample neurons. 
		Here, we use the model parameters $ \beta = 0.5 $, $ \theta = 2 $, $ \Sigma_v = 2 \cdot \mathbb{I}^{2 \times 2}  $, $ g_0 = 80 $, $ \Sigma_d^{1/2} =   \sigma_d \cdot \mathbb{I}^{2 \times 2} $ with $ \sigma_d = 0.2 $, and the place field centers $ \vecmu_d $ are distributed on an equally-spaced grid in the spatial domain covering the whole box $ \D $.
		The trajectory shown here was run for $ T = 100 \cdot \theta^{-1} $.
		\textbf{(b)} Sample spike train corresponding to the first 20 time constants $ \theta^{-1} $.
		\textbf{(c)} Decoding the spike train in (b) with the sNPF, BPF and the ADF.
		Note that the ADF trajectory may leave the box because reflecting boundary conditions cannot be incorporated.
		\textbf{(d)} Numerical filtering performance in terms of MSE for the three filtering algorithms, averaged over $ T = 1000 \cdot \theta^{-1} $.
		\textbf{(e)} A large drift parameter $ \theta $ is able to `pull' the ADF estimate away from the boundaries, making it almost optimal again. 
		Note that for this example exceptionally large values of $ \theta $ are needed. 
		Here, $ \sigma_d = 0.1 $.
	}
\end{figure}

The process $ \mathbf{k}_t $ is the minimal process needed to restrict the position $ \vecx_t $ to lie within the box $ \mathcal{D} $  \cite{Lepingle1995}.
A possible realization of the process  $ \mathbf{k}_t $ is \cite{Hanks2017}:
\begin{eqnarray}
\mathbf{k}_t & = & \begin{cases}
0 & \text{if }\vecx_t \in \D \\
- \left( \mathbf{n}(\vecx_t)^T \vecv_t \right) \mathbf{n}(\vecx_t) & \text{if }\vecx_t \in \partial \D
\end{cases}, \label{eq:Paper 3 - 2d model - boundary process}
\end{eqnarray}
where $ \vecn $ is a normal vector perpendicular to the boundary $ \partial \D $.
This process effectively cancels the velocity component perpendicular to the boundary and thus implements boundary conditions of the von Neumann type\footnote{i.e.~the probability flux vanishes at the boundary $ \partial \D $ of the box.}.
A sample trajectory of the resulting process (with corresponding responses of spatially tuned neurons) is shown in Figure \ref{fig:Paper3 - 2D place cell decoding - hidden trajectory}.

The boundary condition can be directly implemented in the sNPF and the BPF using a projection approach (cf.~\cite{Lepingle1995,Dangerfield2012,Hanks2017}).
However, in the ADF, these cannot be directly implemented, because the process in Eq.~\eqref{eq:Paper 3 - 2d model - boundary process} imposes a strong nonlinearity, such that the Gaussian approximation is violated at the boundaries.
The only possibility is to avoid explicitly modeling the boundary process $ \mathbf{k}_t $ in the ADF altogether, and rely on the filter estimate being 'driven back' into the box $ \D $ by the spiking events (Figure \ref{fig:Paper3 - 2D place cell decoding - filter trajectory}).

The numerical performance of the sNPF for this rather complex model is again strikingly close the the benchmark (BPF) for all parameters tested, as shown in Figure \ref{fig:Paper3 - 2D place cell decoding - MSE}.
On the other hand, the numerical performance of the ADF strongly depends on the particular choice of the parameters.
More precisely, if the simulation setup is sufficient to keep the filter trajectory well within the boundaries (e.g. for large effective firing rates or for a strong drift towards the center modified by $ \theta $, cf.~Figure \ref{fig:Paper3 - 2D place cell decoding - MSE harder fail}), then the ADF performs reasonably well.
If this is not the case, the numerical performance of the ADF is much worse than that of the sNPF.

\section{On the COD in weighted approaches}

\label{sec:Paper3 - COD}


In contrast to standard particle filtering algorithms, such as the sequential importance sampling algorithm \cite{Doucet2000}, the particles in Eq.~\eqref{eq:Paper 3 - sNPF} are considered equally weighted.
This comes with what we consider a huge advantage: weighted particle filtering approaches are known to suffer from inevitable weight decay over time \cite{Daum2003}, an effect that becomes more severe as the dimensionality of the problem increases \cite{Surace2017} (see also Section S~4.3 in the SI).

To illustrate the problem, consider the BPF \cite{Doucet2000}, which represents the posterior with a weighted empirical distribution:
\begin{eqnarray}
p(\vecx_t|\N_{0:t}) & \approx &  \sum_{i=1}^{P} w_t^{(i)} \delta(\vecx_t - \vecx_t^{(i)} ), \label{eq:Paper 3 - sNPF - PF empirical density}
\end{eqnarray}
where the (normalized) importance weights $ w_t^{(i)} $ change dynamically over time.

\begin{figure}
	\centering
	\begin{subfigure}{0.44\textwidth}
		\caption{ \label{fig:Paper 3 - COD Peff}
		}
		\includegraphics[width=\textwidth]{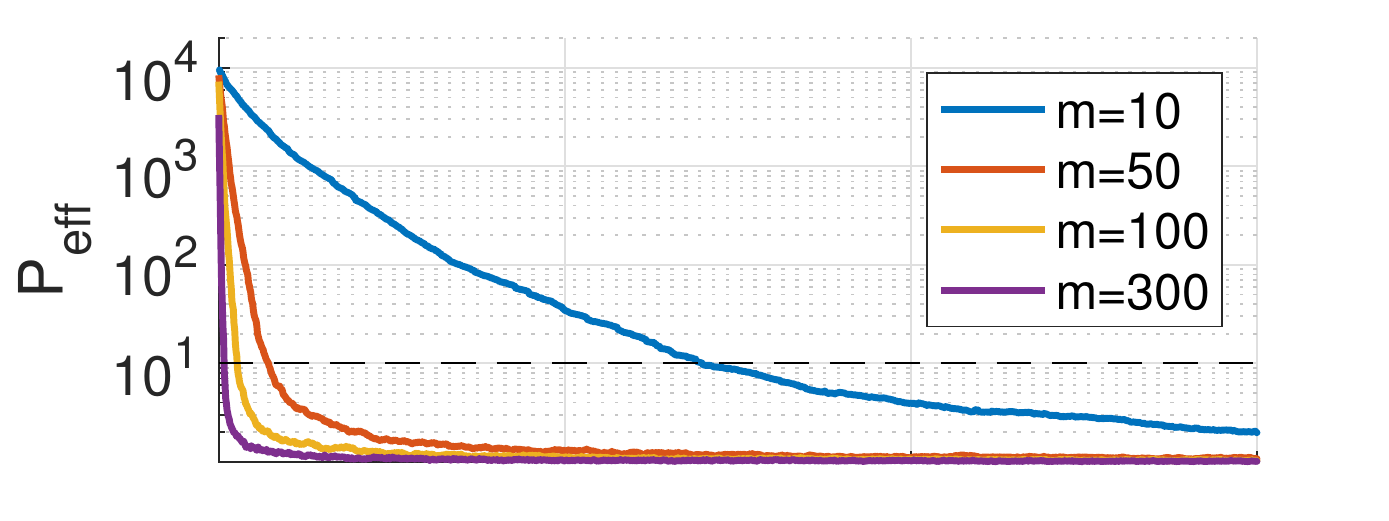}
		\\
		\includegraphics[width=\textwidth]{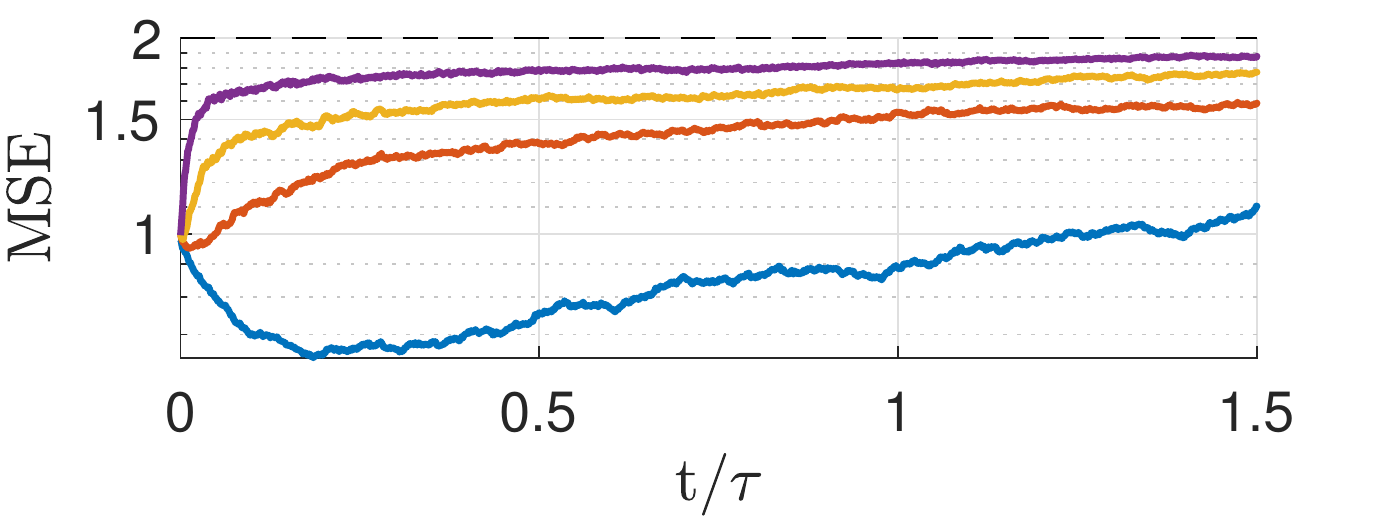}
	\end{subfigure}
	\begin{subfigure}{0.54\textwidth}
		\caption{ \label{fig:Paper 3 - COD stoppingtimes}
		}
		\includegraphics[width=\textwidth]{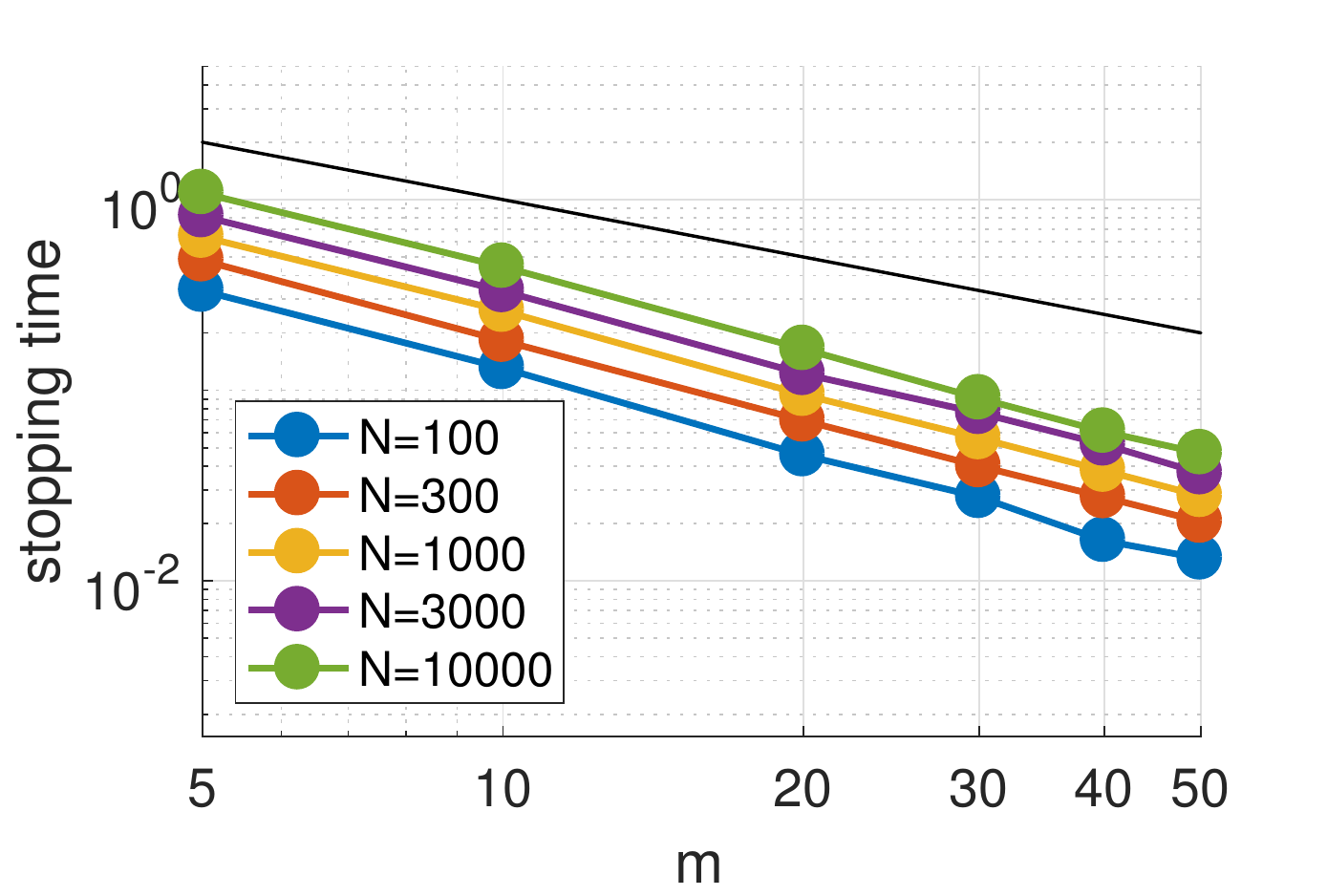}
	\end{subfigure}
	\\
	\begin{subfigure}{0.5\textwidth}
		\caption{ \label{fig:Paper 3 - COD scaling}
		}
		\includegraphics[width=\textwidth]{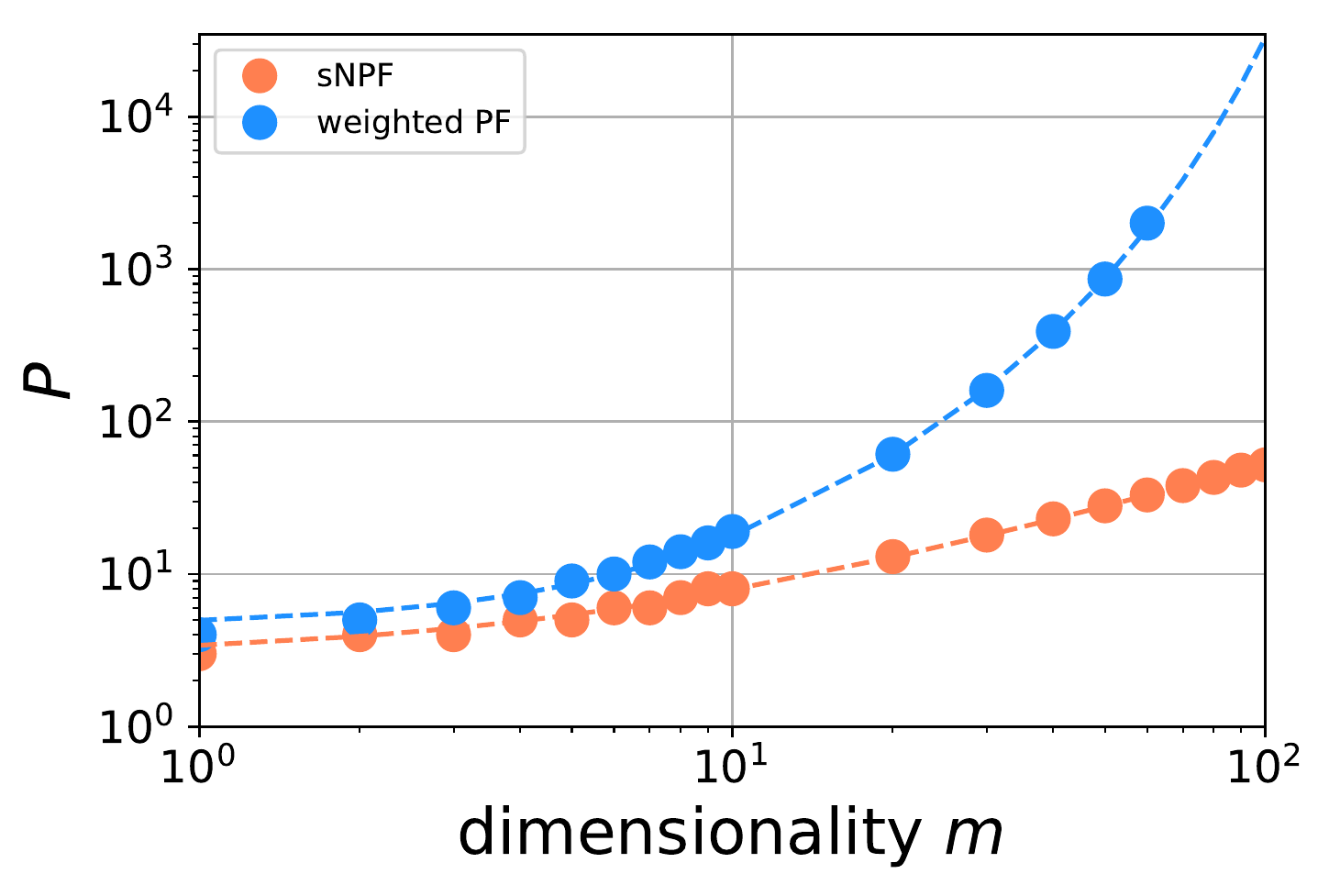}
	\end{subfigure}
	\caption[]{\label{fig:Paper 3 - COD}
		{\bf The COD in weighted particle filters.}
		{\bf{(a)}} 
		Importance weights of the BPF decay over time, illustrated by the depletion of the effective number of particles $ P_\text{eff} $.
		At the same time, the MSE increases, indicating a significant drop in performance.
		Note that $ \text{MSE} = 2 $ denotes a filtering performance that corresponds to taking a single sample from the hidden dynamics in Eq.~\eqref{eq:Paper3 - Problem dx}, employing no knowledge of the observations.
		{\bf{(b)}}
		Time to reach $ P_\text{eff} = 10 $ (`stopping time') as a proxy for the particle decay time scale.
		The black line denotes a $ 1/m $ scaling.
		The time to reach $ P_\text{eff} = 10 $ decreases faster upon increasing the problem dimensionality $ m $.
		In (a) and (b), the model is a multidimensional Ornstein-Uhlenbeck (OU) process with exponential firing rate function.
		{\bf{(c)}}
		The number of particles needed for a fixed performance (of $ 80 \% $ of the optimal MSE) across dimensions scales exponentially in a BPF and linearly in the sNPF. 
		In this simulation, a nonlinear hidden dynamics, $ f(x) \propto x(x^2-1) $, with exponential firing rate function $ g(x) \propto \exp \left(  2 x \right) $ was used.
	}
\end{figure}

For continuous-time models, it is possible to write down an SDE that describes the dynamics of the weights (see Section S~4.2.1, Eq.~S-43, in the SI).
From this, we obtain an SDE for the inverse of the `number of effective particles' $ \Peff^{-1} = (\sum_i (w_t^{(i)})^2 )^{-1} $, which may be considered as an empirical estimate of how many particles are effectively representing the posterior.
If all weights are equal, i.e. $ w_t^{(i)} = 1/P \  \forall i $, then $ P_\text{eff} = P $.
If, on the other hand, one weight is $ 1 $ and all the other weights are zero, we find $ P_\text{eff} = 1 $.
In the latter case, the system is highly degenerate, and only a single particle contributes to representing the posterior.
The SDE for $ P_\text{eff}^{-1} $ reads (Eq.~S-45 in the SI):
\begin{eqnarray}
d P_\text{eff}^{-1}  & = &  \sum_{i,d=1}^{P,m} (w_t^{(i)})^2 \left[ -2 (g_d(\vecx_t^{(i)} ) - \bar{g}_d)\,dt + (g_d(\vecx_t^{(i)})^2 -\bar{g}_d^2) \frac{1}{\bar{g}_d^2} \, dN_t^{(d)} \right],
\end{eqnarray}
where the bar denotes an empirical estimate according to the particle distribution.
In Figure \ref{fig:Paper 3 - COD Peff}, we perform simulations showing that the effective number of particles $ \Peff $ decreases over time (Figure \ref{fig:Paper 3 - COD Peff}).
This is accompanied by an increase in MSE that tends towards $ \text{MSE} = 2 $, corresponding to that of an independent draw from the hidden process, i.e.~without any knowledge of the observations.

Let us now make this numerical observation of a dynamical weight degeneration more formal and consider initializing a system with equal weights and particle positions that are drawn from the true posterior.
Similarly to the argument made in Surace et al.~\cite{Surace2017}, this equation gives us an estimate of the initial rate of decay of the effective number of particles: 
since this equation has $ m $ summands, a rough estimate of the decay time scale of $ \Peff $  is $ \tau_{decay} \propto \frac{1}{m} $ (please see Section S~4.3 in the SI for more details on this argument).
Even more, the simulations in Figure \ref{fig:Paper 3 - COD stoppingtimes} suggest that this analytical argument seems to even \emph{underestimate} the rate on which the particles degenerate, as it seems to happen - at least numerically - on an even faster time scale.

Thus, the higher the dimensionality of the system, or rather the dimensionality of the observed process, the more particles need to be in the system to compensate for the fast decay - even if numerical tricks such as particle resampling are applied\footnote{In Surace et al.~\cite{Surace2017} it is demonstrated that resampling improves on a timescale that is independent of the dimension, and thus is ineffective for high-dimensional systems.}.
To demonstrate this, we performed simulations where we determined the minimum number of particles that are needed for a fixed numerical filtering performance of $ 0.8 \% $ of the optimal MSE\footnote{Which can be determined for this model by determining the optimal MSE in 1D with a BPF with a sufficiently large number of particles. We exploited the fact that, if $ \vecf(\vecx_{t}) $ and $ \vecg_t(\vecx_{t}) $ are independent in each dimension, then $ \text{MSE}_{\text{opt},mD} =  m\text{MSE}_{\text{opt},1D}$.} in a nonlinear model, both with the sNPF and with a standard BPF.
The minimum number of particles needed for the BPF increases exponentially with problem dimensionality $ m $, whereas the BPF only scales linearly with dimensions (Figure \ref{fig:Paper 3 - COD scaling}).

In earlier work, Huang et al.~\cite{Huang2009} suggested particle filters as a possible alternative for decoding of nonlinear models, but also mention numerical issues found earlier \cite{Ergun2007}.
Practically, our analysis aims to demonstrate that weighted PFs are not advantageous for decoding if the dimensionality of the observations, for instance the number of recording electrodes, is large.
Thus, for these cases, unweighted particle filtering approaches offer a suitable alternative to weighted approaches.

%
%

\section{Adaptive Encoding} \label{sec:Adaptive Encoding}

In a broad sense, encoding aims to find the relationship between the hidden stimulus $ \vecx_t $ and the neural spike train.
Here, we narrow this task down to finding the parameters of a previously specified model.
In this section, we use the particle filter ansatz outlined in the previous section to approximate an encoding scheme.
The resulting algorithm is suitable for both offline and online (`adaptive') encoding when the hidden state cannot be directly observed, for instance when adjusting parameters in brain-machine interfaces \cite{Donoghue2002,Shanechi2016}.

\subsection{Maximum likelihood encoding}

As an objective function for learning the model parameters in Eq.~\eqref{eq:Paper3 - Problem dN} and \eqref{eq:Paper3 - Problem dx} we use the incomplete-data log likelihood, which is given by (see Section S~5.1 in SI):
\begin{eqnarray}
\Lambda_t^\theta & = & \sum_{d = 0}^m  \left( \int_{0}^{t} \log\frac{ \E \left[ g_{d} (\vecx_{s}) | \N_{0:s} \right]  }{\lambda_0} dN_{d,s} +  \int_{0}^{t} (\lambda_0 - \E \left[ g_{d} (\vecx_{s}) | \N_{0:s} \right] ) \, ds \right), \label{eq:Paper3 - log likelihood PP}
\end{eqnarray}
where $ \lambda_0 $ is a constant reference firing rate, which is needed in the derivation and does not affect likelihood optimization.
Since in a general task we do not have access to the posterior expectations, these are approximated with $ \E \left[ g_{d} (\vecx_{t}) | \N_{0:t} \right]  \approx \ev{ g_{d,t} } $ from the sNPF particle estimates according to Eq.~\eqref{eq:Paper 3 - sNPF particle representation}.
From this perspective, the optimal set of parameters $ \theta $ maximizes the log likelihood function for a given set of observations $ \N_{0:t} $.

\subsection{Learning rules}

Update rules for the parameters $ \theta $ can be obtained by performing a gradient ascent on Eq.~\eqref{eq:Paper3 - log likelihood PP}, using sNPF approximations.
\begin{eqnarray}
d \theta  & = & \eta_\theta \, \partial_\theta \Lambda_t^\theta \nonumber
\\
& \approx & \eta_\theta \, \sum_{d=0}^m \int_0^t  \left( \partial_\theta \ev{g_{d,s}} \right) \left( \frac{1}{ \ev{g_{d,s}}} dN_{d,s} - ds \right) \nonumber
\\
& = &  \eta_\theta \int_0^t  \left( \partial_\theta \ev{\vecg_s}  \right)^T \text{diag}(  \ev{\vecg_{s} } )^{-1} \left( d\N_{s} - \ev{\vecg_s} ds \right) 
\end{eqnarray}
where $ \eta_\theta $ is a (possibly time-dependent) learning rate.
Noting that $ \ev{\vecg_t} $ can be considered a posterior estimate of the firing rate variance $ \Sigma_{g,t} $, the learning rules read:
\begin{eqnarray}
d \theta & = &  \eta_\theta \int_0^t  \left( \partial_\theta \ev{\vecg_s}  \right)^T \Sigma_{g,t}^{-1} \left( d\N_{s} - \ev{\vecg_s} ds \right)  \label{eq:Paper3 - learning rules offline}
\end{eqnarray}

Online learning rules are obtained by maximizing the infinitesimal \emph{increments} of the log likelihood in Eq.~\eqref{eq:Paper3 - log likelihood PP} at each point in time, while the sequence of observations $ \N_{0:t} $ keeps flowing in.
This can only be expected to converge if certain ergodicity requirements of the whole stochastic process are fulfilled (a proof for this, albeit for continuous-time observations, can be found in \cite{Surace2016}).
Assuming process ergodicity, the online learning rules read:
\begin{eqnarray}
d \theta  & = & \eta_\theta \left( \partial_\theta \ev{\vecg_t}  \right)^T \Sigma_{g,t}^{-1} \left( d\N_{t} - \ev{\vecg_t} dt \right) \label{eq:Paper3 - learning rules online}
\end{eqnarray}

Note that for both the offline and online learning rules depend on the derivative of a posterior expectation, i.e.~$ \partial_\theta \ev{ \vecg_{t} } $.
When computing this derivative, we have to take into account the change of the rate function $ \vecg $ as well as the infinitesimal changes in the filter with respect to the parameters $ \theta $.
The latter is referred to as 'filter derivative' and in general cannot be computed in closed form.
However, using the particle representation of the sNPF allows us to compute a filter derivative for every single particle and thus approximate Eq.~\eqref{eq:Paper3 - learning rules offline} and \eqref{eq:Paper3 - learning rules online}.
For details, see Section S~5.2 in the SI.

\subsection{Learning the gain matrix $ W_t $}

As an alternative to using the EKSPF, we can determine the gain matrix by treating it as a parameter (`decoding parameter') that can be learned online.
We can immediately write down the online learning rule for the components of the gain matrix, $ W_{ij} $ using Eq.~\eqref{eq:Paper3 - learning rules online} and the fact that there is no explicit dependence of the functions $ \vecf $ and $ \vecg $ on $ W $ (cf.~Eq.~S-50 in SI  Section S~5.2):
\begin{eqnarray}
d W_{ij}  & = & \eta_{W_{ij}} \left( \ev{ D\vecg_t \cdot \vecalph_{W_{ij},t}  } \right)^T \Sigma_{g,t}^{-1} \left( d\N_{t} - \ev{\vecg_t} dt \right),
\end{eqnarray}
with filter derivative dynamics
\begin{eqnarray}
d  \vecalph_{W_{ij},t}^{(k)}  & = & D\vecf_t^{(k)} \cdot \vecalph_{W_{ij},t}^{(k)}  dt \nonumber \\
& & + \left( d\mathbf{N}_t - \mathbf{g}_t^{(k)} \,dt \right)_j  \mathbf{e}_i-  W_t \cdot   D\vecg_t^{(k)} \cdot \vecalph_{W_{ij},t}^{(k)} \,dt,
\end{eqnarray}
where $  \mathbf{e}_i $ denotes the unit vector in $ i $-direction, and  $ \left( D\vecf_t \right)_{i,j} = \partial_{x_j} f_i$ and $ \left( D\vecg_t \right)_{i,j} = \partial_{x_j} g_i$ denotes the Jacobian of the functions $ \vecf_t $ and $ \vecg_t $, respectively.

Note that the resulting filtering scheme is based on the heuristic ansatz in Eq.~\eqref{eq:Paper 3 - sNPF} that is compensated for in a rigorous way by choosing the gain such that  the incomplete-data log likelihood is maximized.


\section{Discussion}

In this paper, we proposed an ansatz for a recursive filtering algorithm for point-process observations, the spike-based Neural Particle Filter (sNPF).
Since it relies on an approximation of the posterior density by equally weighted, empirical samples, it can be seen as an unweighted particle filtering method.
One realization of this general ansatz uses an empirical approximation for the weight matrix that governs the strength of the particle updates due to the point-process observations, has been put forward before and is called ensemble Kushner-Stratonovich-Poisson filter (EKSPF,\cite{Venugopal2015}).
Despite its relevance for neuroscience and in particular decoding, this filter has not received much attention in the community yet.
In addition, a thorough analysis of its capabilities is hitherto lacking.
Here, we showed that its decoding performance in a toy decoding task, which was motivated by a realistic problem setting, is almost optimal, and even may outperform an ADF due of its flexibility.
Further, we demonstrated its usefulness for higher dimensional models, which become essential in times when growingly sophisticated probes allow for an ever increasing amount of simultaneously recorded neurons.
In particular, we showed that it is a promising candidate for a filtering algorithm that avoids the COD, in contrast to standard PFs.
Lastly, the sNPF can be used to approximate a likelihood function, that allows the derivation of both online and offline learning rules for system identification.

The idea to frame the decoding task as a question of Bayesian inference is certainly not a new one and has been around ever since the discovery that neuronal spiking can be, to some extent, modeled by a Poisson process.
For static stimuli, maximum \emph{a-posteriori} (MAP) methods have proven successful, in particular for generalized linear models where the (complete-data) likelihood function is convex \cite{Pillow2011,Macke2011}.
A straightforward extension to dynamic stimuli can be obtained by considering averaging spikes over a fixed time window \cite{Zhang1998,Pillow2011}.
A similar problem as with ML decoding arises here as well (Figure \ref{fig:Paper3 - 1D place cell decoding - sNPF vs ML}):
the decoding accuracy in a dynamic setting will crucially depend on how large the time window is chosen.
Further, MAP estimates tend to exhibit a large decoding error whenever the posterior density deviates significantly from a unimodal density \cite{Ahmadian2011}.

Alternatively, variational inference (cf.~\cite{Opper2009}) aims at matching the posterior and a (usually Gaussian) proposal by minimizing the Kullback-Leibler divergence between the two densities with respect to the proposal parameters.
For specific models with point-process observations, the parameters can be solved for in closed form \cite{Macke2011}, but it is unclear how this result would relate to other models, e.g.~Gaussian-shaped tuning curves.
Further, the discussion in \cite{Macke2011} is restricted to a static setting, and it would be interesting to see whether it can be extended to a dynamical setting, which has been done for instance for continuous-time observations \cite{Archambeau2008}.

An important class of models used for decoding are (Gaussian) assumed density filters (ADF), which are particularly suited to decode from Gaussian-shaped tuning curves.
ADFs are available for both discrete \cite{Brown1998,Eden2004} and continuous time \cite{EdenBrown2008,Harel2016} models.
The Gaussian approximation in the filters by Brown, Eden and coworkers result from a second-order Taylor expansion of the posterior density, and is thus roughly along the lines of an extended Kalman filter (EKF) for continuous-time observations.
As a second example, the ADF in \cite{Harel2016} directly approximate posterior estimates in the filtering equations for mean and variance under the Gaussian assumption.
In this regard, it is surprising that, even if the hidden dynamics is nonlinear (cf. Section \ref{sec:Paper 3 - One-dimensional toy model}), the ADF exhibits almost optimal performance - as long as the posterior is still sufficiently Gaussian-shaped.
For our simulations, we chose the ADF by Harel et al.~\cite{Harel2016} as a comparison, because the mindset in the heuristic derivation of the sNPF is similarly circular:
assuming the posterior to be approximated by $ \delta $-masses, we approximate any posterior estimates that appear in the dynamical equations of the particles by estimates under this empirical density.

Both ADF versions suffer from the fact that they have to be computed explicitly for each particular model, which can turn out to be tedious or even impossible analytically.
Thus, they are also very limited in their model flexibility:
as long as the decoding task is comparatively simple with sufficiently linear hidden dynamics and exponential rate function, this class of filters offers a suitable and computationally efficient decoding scheme.
However, more complex tasks, for instance when introducing boundary conditions like in in Section \ref{sec:Paper 3 - 2D RBM model}, first might imply a serious violation of the Gaussian assumption at the boundaries and, second, might lead to a significant drop in performance of the ADF when neglected.
For these `complex' tasks, the sNPF turns out to be superior, because it can accommodate a huge variety of generative models, linear and nonlinear.
Due to the specific representation scheme, it is possible to represent multimodal posterior densities.
Further, the implementation does not require computing any further instances of the model, such as posterior expectations or derivatives, but just needs being able to evaluate the hidden function $ \vecf $ and the rate function $ \vecg $ at the particle positions.

Since it approximates the posterior density with empirical samples, the sNPF can be seen as a PF, albeit without importance weights.
Standard weighted PFs are still considered state-of-the-art filtering algorithms, because they are asymptotically exact in the limit of infinite particles, can be derived rigorously from a change of measure approach (cf.~Section S~4 in the SI) and, last but not least, are easily implementable numerically.
Because they are not limited in their flexibility, particle-based methods may enhance decoding performance by tackling multi-modality \cite{Ahmadian2011} and/or model nonlinearities \cite{Huang2009}.
Unfortunately, it is known (cf.~also Section \ref{sec:Paper3 - COD}) that they suffer from an inevitable weight decay over time, leading to a decline in the number of particles that effectively represent the posterior density.

The time scale on which weight decay occurs decreases with the number of observable dimension, effectively leading to a COD, as shown in Section \ref{sec:Paper3 - COD}.
In continuous-time models, it might help to choose better proposal densities, which already take into account the observations in the proposal and then propagate the particles into significant regions in space. 
Finding such a proposal for point process observations is an open problem even in discrete time, as for instance the `optimal proposal' \cite{Doucet2000} can be computed only for very specific models.
Further, resampling the particles frequently is known to numerically boost the BPF performance.
However, since the time scale on which resampling is beneficial is independent of the problem dimensionality (cf.~\cite{Surace2017})\footnote{There is no reason to believe that this should be not the case for point-process observations as well.}, resampling is not a sufficient technique to avoid the COD.
Another remedy could be to use an ADF as proposal density for the particles \cite{Ergun2007}, but it is has not been studied yet whether this approach can avoid or mitigate the COD.
Unweighted PF approaches, such as the Neural Particle filter (NPF, \cite{Kutschireiter2017} or the Feedback Particle filter (FBPF, \cite{Yang2013}), hold the promise of avoiding the COD (Figure \ref{fig:Paper 3 - COD}, cf.~also \cite{Surace2017}), because they trivially do not suffer from weight decay.
The derivation of the sNPF relies on a heuristic similar to related equally-weighted particle filtering approaches for continuous-time observations (e.g.~\cite{Kutschireiter2017}), and as such it is not surprising that similar properties emerge.

Here, we have used the sNPF to approximate the incomplete-data log likelihood, i.e.~the likelihood function of the spike train without knowledge of the hidden state.
The resulting learning rules can thus be used for an unsupervised learning task, which has not received much attention in neuroscience yet.
Usually, encoding is done in a supervised way, because one has access to the ground truth.
One of the few examples of parameter learning, that is both online and unsupervised, was tackled in \cite{Eden2004} by state augmentation: 
imposing a linear dynamics on the parameters, the same algorithm is applied for the learning task as for the decoding task.
In general, it is not clear whether this algorithm maximizes the log likelihood of the whole observation sequence, in particular when the filtering algorithm is approximate.
For learning with the sNPF, a similar question arises:
how can we be sure that learning is accurate if the filter is approximate?
Fortunately, it can be shown (for continuous-time models) that parameter learning based on maximum likelihood can make up for an approximate filter, such that estimated state and estimated parameters together maximize the incomplete-data log likelihood \cite{Surace2016}.
Offline learning can in principle be tackled with huge flexibility by expectation maximization (EM, \cite{Dempster1977}) combined with particle smoothing \cite{Kantas2012a}, which can be computationally expensive.
Alternatively, for models where it is applicable if applicable, a Gaussian approximation leads to closed-form EM update equations \cite{Macke2011}.

A serious limitation of the sNPF is that it apparently does not capture the full posterior density.
By construction, the particle dynamics are such that the dynamics of the first posterior moment matches that from the first moment of the particle density, under the assumption that higher-order moments in this particle dynamics are matched (in the case of the EKSPF).
It is surprising that, despite this assumption to be violated (similarly to an ADF), it does not seem to significantly affect state estimation as measured by MSE.
However, already the variance estimated from the particle distribution deviates to a large extent from that estimated by a BPF.
Analytically, we find that it does not take into account the instantaneous reduction in variance following an event.
Further, the variance in generally evolves slower than that of the benchmark.
To tackle this particular problem, one would need to tune the whole ansatz (not just the gain matrix), for instance by choosing a different noise covariance, such that, at least in between events, the variance can be matched.
In this regard, a thorough error analysis for the EKSPF, which is hitherto lacking, might help to choose the model accordingly, if one is interested in accurately determining higher-order posterior moments.
In general, it would be desirable to ultimately derive an unweighted particle filter that is asymptotically exact, such as the FBPF \cite{Yang2013} for continuous-time observations.


Despite its limitations, the sNPF is a decoding algorithm that is suitable for spike-train decoding.
In particular, it becomes an interesting alternative to an ADF for complex models, which exhibit strong nonlinearities, boundary conditions, or cannot be computed with an ADF in closed form.
For high-dimensional models, since it does not suffer from the COD, it outperforms a BPF for a limited number of particles.
Thus, the sNPF might prove beneficial when compared to standard approaches as more and more neurons can be simultaneously recorded.
Lastly, it might motivate related, similarly scalable algorithms, that will become crucial in the future of neuroscience.

%

%



\bibliographystyle{vancouver} 
\bibliography{library}      

\end{document}


\maketitle

\label{sec:Paper3 - Appendix}
\section{Filtering with point-process observations}

\subsection{The generative model}

In neuroscience, one often has to deal with discrete events such as spikes.
These are often modeled as point processes, or, more specifically, as Poisson processes with time-varying and possibly history-dependent rate.
Often, spike trains are modeled as doubly-stochastic processes, where the rate is again a stochastic process, which mathematically corresponds to a Cox process \cite{Cox1955}.

The filtering problem with point process observations can be formulated as a generative model.
The task is to infer the hidden state $ \vecx_{t} $, which evolves according to an It\^{o} SDE, from the history of $ m $ counting processes $ \vecN_{0:t} $, with $ \vecN_t \in \mathbb{N}_{0^+}^m $.
The discrete increments of the counting processes are drawn from an $ m $-dimensional Poisson distribution:
\begin{eqnarray}
d \vecx_t & = &  \vecf(\vecx_t,t) \, dt + \Sigma_x^{1/2}(\vecx_t,t) d \vecw_t,  \label{eq:GM - hidden}\\
d \vecN_t & \sim & \text{Poisson} \left( \vecg(\vecx_t,t) \, dt \right), \label{eq:GM - dN}
\end{eqnarray}
where $  \vecg: \mathbb{R}^n \times \mathbb{R} \to \R^m_{>0} $ is the instantaneous rate function (sometimes referred to as `intensity function') of the Poisson process.
In principle, the intensity function may have a dependence on the history of the counting process.
Since it does not alter the treatment nor the solutions to the filtering problem \cite{Surace2015}, we will leave this out as an explicit argument in the intensity function for brevity.

\subsection{The formal solution} \label{sec:The-formal-solution} 

Similarly to the standard filtering problem with continuous-time observations (cf.~\cite{Bain2009}), the solution to the filtering problem can be derived using the\emph{ change of probability measure} method.
We will briefly outline the derivation, but refer the reader to the very detailed derivation offered in \cite[p.~170 ff.]{bremaud1981point} or to the more intuitive derivations in \cite[SI]{Bobrowski2009} and \cite[p.~41ff]{Surace2015}.

A full probabilistic solution to the filtering problem is known if we can compute the posterior expectation (if it exists) of any real-valued measurable function $ \phi $,
\begin{eqnarray}
\E_\P \left[ \phi(X_t) | \N_{0:t} \right] & = &\int_{-\infty}^\infty p (\vecx_t|\N_{0:t}) \phi(\vecx_t) \, d \vecx_t, \label{eq:FS - posterior expectation definition}
\end{eqnarray}
where we use subscript $ \P $ to indicate expectations with respect to the posterior measure.
This equation can be rewritten in terms of the expectation with respect to a reference measure $ \Q $, using the Kallianpur-Striebel formula:
\begin{eqnarray}
\E_\P \left[ \phi_t | \N_{0:t} \right] & = & \frac{\E_\Q [\phi_t\, L_t | \N_{0:t} ]}{\E_\Q [ L_t | \N_{0:t} ]} = \frac{1}{Z_t} \expect{\phi_t L_t }_\Q, \label{eq:FS - KallianpurStriebel}
\end{eqnarray}
where $ L_t := \frac{dP_t}{dQ_t} $ denotes the Radon-Nikodym derivative, $ Z_t := \E_\Q [ L_t | \N_{0:t} ]$ denotes a time-dependent normalization, and $ \expect{\phi_t L_t }_\Q := \E_\Q [\phi_t\, L_t | \N_{0:t} ]  $ was introduced to keep the notation concise.
The estimation problem under $ \P $ is thus replaced by an estimation problem under $ \Q $, which can be chosen freely provided it is equivalent to $\P$.

A convenient choice for $ \Q $ is a measure under which the point processes become independent from the signal process $ \vecx_t $, while the signal process itself remains unaffected.
Using a Girsanov-type theorem (see \cite{bremaud1981point}, Chapter 2, Theorems T2 and T3), we may choose an equivalent measure $\Q$ under which all point process have a constant rate $\lambda_0$.
The Radon-Nikodym derivative reads
\begin{eqnarray}
L_t & = & \prod_{d = 0}^m \left( \prod_{0 <s \leq t} \frac{g_d (\vecx_t) }{\lambda_0} \Delta N_{d,s}   \right)  \exp \left( \int_{0}^{t} (\lambda_0 - g_d (\vecx_t)  ) \, ds \right). \label{eq:FS - RadonNikodym}
\end{eqnarray}
Under $\Q$ the stochastic differential can be taken inside the expectation (see \cite{bremaud1981point}, Chapter 3, Theorem T7), and we therefore obtain
\begin{eqnarray}
d\E_\P \left[ \phi_t | \N_{0:t} \right] & = & d \left( \frac{1}{Z_t} \expect{\phi_t L_t }_\Q \right) \nonumber 
\\
& = & \frac{1}{Z_t} \, d \expect{\phi_t L_t }_\Q + \expect{\phi_t L_t }_\Q \, d\frac{1}{Z_t} + (d \expect{\phi_t L_t }_\Q) \, (d\frac{1}{Z_t}) \nonumber
\\
& = & \frac{1}{Z_t} \, \expect{ d( \phi_t L_t ) }_\Q + \expect{\phi_t L_t }_\Q \, d\frac{1}{Z_t} +  \expect{d(\phi_t L_t) }_\Q \, (d\frac{1}{Z_t}). \label{eq:FS - posterior expectation evolution general}
\end{eqnarray}
From Eq.~\eqref{eq:FS - RadonNikodym} we obtain the evolution equation of the Radon-Nikodym derivative, which reads:
\begin{eqnarray}
dL_t & = & L_t \cdot \sum_{d = 1}^{m} \left( \frac{g_d(\vecx_t)}{\lambda_0}  -1 \right) (dN_{d,t} - \lambda_0 dt). \label{eq:FS - RadonNikodym evolution}
\end{eqnarray}
With this evolution equation we can compute $ \expect{ d(\phi_t L_t) }_\Q  $ in Eq.~\eqref{eq:FS - posterior expectation evolution general}:
\begin{eqnarray}
\expect{ d(\phi_t L_t) }_\Q & = & \expect{ (d\phi_t) \, L_t + \phi_t \, (dL_t) + (d\phi_t) \, (dL_t) }_\Q.
\end{eqnarray}
Using It\^{o}'s lemma to compute the evolution equation for $ \phi_t $ and noting that under Q, the posterior expectation of the Wiener process $ \vecw_t $ is zero due to the independence of signal and observation process, we find:
\begin{eqnarray}
\expect{ d(\phi_t L_t) }_\Q & = & \expect{ \FPO{\phi_t}  \, L_t}_\Q \nonumber \\
& & + \expect{\phi_t  L_t \cdot \sum_{d = 1}^{m} \left( \frac{g_g(\vecx_t)}{\lambda_0}  -1 \right) (dN_{d,t} - \lambda_0 dt) }_\Q. \label{eq:FS - unnormalized expectation evolution}
\end{eqnarray}
The differential $ d \frac{1}{Z_t} $ can be obtained by It\^{o}'s lemma for point processes, using Eq.~\eqref{eq:FS - unnormalized expectation evolution} with $ \phi_t = 1 $.
By plugging Eq.~\eqref{eq:FS - unnormalized expectation evolution} and $ d \frac{1}{Z_t} $ into Eq.~\eqref{eq:FS - posterior expectation evolution general} and rewriting everything in terms of expectations under $\P$ using Eq.~\eqref{eq:FS - KallianpurStriebel}, we finally obtain the evolution equation for the filter:
\begin{eqnarray}
d\expect{\phi_t}_\P &:=& d\E_\P \left[ \phi_t | \N_{0:t} \right] \nonumber
\\
& = & \expect{\FPO{\phi_t}}_\P \, dt + \sum_{d=1}^{m}  \frac{\text{Cov}_\P(g_d , \phi_t)}{\expect{g_d}_\P}  \left(dN_{d,t} - \expect{g_d}_\P \,dt \right). \label{eq:FS - PostExpect evolution solution}
\end{eqnarray}
Equation \eqref{eq:FS - PostExpect evolution solution} is a \emph{formal} solution, because it suffers from a closure problem similarly to that of standard filtering problems.
It becomes apparent when attempting to actually compute Eq.~\eqref{eq:FS - PostExpect evolution solution}: 
here, one needs access to the posterior covariance between the rate functions $ g_d $ and the function $ \phi $, which generally relies on higher-order moments of the posterior distribution.
Thus for practical use, Eq.~\eqref{eq:FS - PostExpect evolution solution} needs to be approximated using a finite-dimensional representation.

\section{Approximate closed-form decoding algorithms}

\subsection{Static decoding algorithms: ML decoder and MAP decoder \cite{dayan2001theoretical}}
\label{sec:Paper3 - Appendix - ML decoder}

Here we will derive a simple maximum likelihood (ML) decoder, following \cite[p.~101ff]{dayan2001theoretical}.
In ML decoding, the decoded stimulus $ \hat{\vecx} $ is the one that maximizes the likelihood of the observations, in this case the spike train $ \vecs $, such that 
\begin{eqnarray}
\hat{\vecx} = \underset{x}{\text{argmax}}\  p(\vecs|\vecx) = \underset{x}{\text{argmax}}\  \log p(\vecs|\vecx)
\end{eqnarray}
The spike train, or count, $ \vecs $ is generated from the Poisson process in Eq.~\eqref{eq:Paper3 - Problem dN}, but assumes a finite time interval $ \Delta $:
\begin{eqnarray}
p(\vecs|\vecx) & = & \sum_d \text{Poisson} (s_d ; \, g_d(\vecx) \Delta ),
\end{eqnarray}
where $ g_d $ and $ s_d $ denote the instantaneous firing rate and spike count, respectively, of neuron $ d $ according to Eq.~\eqref{eq:Paper 3 - g_d(x)}.

Hence,
\begin{eqnarray}
\log p(\vecs | \vecx) & = & \sum_d \text{Poisson} (s_d ; \, g_d(\vecx) \Delta ) \nonumber
\\
& = & \sum_s \left[ s_d \log (g_d(\vecx) \Delta) - g_d(\vecx) T - \log (s_d!) \right] \nonumber
\\
& \approx & -\frac{1}{2} \sum_d s_d (\vecx - \vecmu_d)^T \Sigma_d^{-1} (\vecx - \vecmu_d ) + \text{const}.
\end{eqnarray}
Here, we assumed the coverage of the place fields to be sufficiently \emph{dense}, such that $ \sum_d g_d(\vecx)  \approx \text{const} $.
Minimizing the last equation with respect to $ \vecx $ gives the ML estimate:
\begin{eqnarray}
\hat{\vecx} & = & \left( \sum_d s_d \Sigma_d^{-1} \right)^{-1} \left( \sum_d s_d \Sigma_d^{-1} \vecmu_d \right). \label{eq:Paper 3 - Appendix - ML}
\end{eqnarray}

In a similar way, by incorporating prior knowledge $ p(\vecx) $ about the stimulus, a maximum \emph{a-posteriori} (MAP) estimator can be derived:
\begin{eqnarray}
\hat{\vecx} &=& \underset{x}{\text{argmax}}\  \log p(\vecs|\vecx)p(\vecx) . 
\end{eqnarray}
Assuming for instance that the prior density is a Gaussian with mean $ \vecmu_x $ and variance $ \Sigma_x $, we find
\begin{eqnarray}
\hat{\vecx} & = & \left( \sum_d s_d \Sigma_d^{-1} + \Sigma_x ^{-1} \right)^{-1} \left( \sum_d s_d \Sigma_d^{-1} \vecmu_d + \Sigma_x^{-1} \vecmu_x \right). \label{eq:Paper 3 - Appendix - MAP}
\end{eqnarray}

The derivation of Eq.~\eqref{eq:Paper 3 - Appendix - ML} and \eqref{eq:Paper 3 - Appendix - MAP} assumes a static setting, with a static stimulus $ \vecx $ giving rise to a spike count $ \vecs $ in the time interval $ \Delta $.
To generalize these approaches to a non-anticipating, dynamic setting with time-dependent stimulus $ \vecx_t $, one would have to use a sliding window approach, i.e.
\begin{eqnarray}
\vecs_t & = & \int_{t-\Delta}^t d\N_s,
\end{eqnarray}
and then decode $ \vecx_{t} $ from $ \vecs_t $ at each instance in time (as was done for the simulations in Figure \ref{Main-fig:Paper3 - 1D place cell decoding - sNPF vs ML}).
Note that these approaches are agnostic to the formal solution of the filtering problem.

\subsection{The Gaussian ADF \cite{Harel2016}}
\label{sec:Paper3 - Appendix - ADF}

In our simulations in Section \ref{Main-sec:Paper 3 - Decoding position from place cell activity} (main manuscript), we use a Gaussian ADF \cite{Harel2015,Harel2016} in order to compare its performance to that of the sNPF.
In Gaussian ADFs, posterior moments are propagated under the assumption of a Gaussian posterior, effectively leading to update equations of the posterior mean $ \boldsymbol{\mu}_t $ and variance $ \Sigma_t $, which can be derived from Eq.~\eqref{eq:FS - PostExpect evolution solution}.

For a model with Gaussian-shaped tuning curves, the ADF takes the form (cf.~Eq.~(10),(11), (19)-(23) in \cite{Harel2016}):
\begin{eqnarray}
d\vecmu_t & = & \ev{\vecf_t} \, dt + \Sigma_t H^T \sum_{d = 1}^m S_t^d (  \vecmu_d - H \vecmu_t) ( dN_{d,t} - \ev{g_d} dt  ), 
\\
d \Sigma_t  & = & d\Sigma_{\pi,t} + d\Sigma_{c,t} + d\Sigma_{N,t},
\end{eqnarray}
with
\begin{eqnarray}
d\Sigma_{\pi,t} & = & \left( \cov(\vecf_t,\vecx_{t}^T) +  \cov(\vecx_{t},\vecf_t^T) + \Sigma_x \right) dt, 
\\
d\Sigma_{c,t} & = & \Sigma_t H^T \sum_{d = 1}^m  \ev{g_d} \left( S_t^d - S_t^d ( H \vecmu_t - \vecmu_d ) ( H \vecmu_t - \vecmu_d )^T S_t^d \right) H \Sigma_t dt 
\\
d \Sigma_{N,t} & = & - \Sigma_t H^T \sum_{d = 1}^m S_t^d H \Sigma_t \, dN_{d,t}. \label{eq:Paper 3 - Appendix - ADF dSigmaN}
\end{eqnarray}
Here,
\begin{eqnarray}
S_t^d & = & (\Sigma_d + H \Sigma_t H^T)^{-1} ,
\\
\ev{g_d} & = & g_0 \sqrt{ \frac{ \det( S_t^d ) }{ \det (\Sigma_d^{-1} ) } } \exp \left( \frac{1}{2} ( \vecmu_d - H \vecmu_t ) S_t^d ( \vecmu_d - H \vecmu_t )^T\right).
\end{eqnarray}
Further, $ H \in \R^{m \times n} $ is a matrix of full row rank that maps the state $ \vecx_t $ to the observation coordinates.

It is not immediately apparent that the resulting density should still lie on a Gaussian manifold.
In particular, the variance update due to the spikes\footnote{The prior update $ d\Sigma_{\pi,t} $ and the continuous update $ d\Sigma_{c,t} $ are not problematic, because they scale with $ dt $}, Eq.~\eqref{eq:Paper 3 - Appendix - ADF dSigmaN}, will always give a \emph{negative} contribution to the variance of order $ \Sigma_t H^T S_t^d H \Sigma_t  $ whenever there is a spike in neuron $ d $.
Luckily, it can quickly be shown that $ \Sigma_t -  \Sigma_t H^T S_t^d H \Sigma_t  $ is always positive definite\footnote{Note that $ H $ is in general not a square matrix, such that $ H^{-1} $ does not exist. 
	However, since it has full row rank, we can use the right inverse $ H_r^-1 = H^T ( H H^T )^{-1}$, such that $ H H_r^{-1} = \mathbb{I}^{m \times m} $}:
\begin{eqnarray}
\Sigma_t -  \Sigma_t H^T S_t^d H \Sigma_t &>& 0 \nonumber
\\
\Leftrightarrow H_r^{-1} -  \Sigma_t H^T S_t^d  &>& 0 \nonumber
\\
\Leftrightarrow H_r^{-1} (\Sigma_d + H \Sigma_t H^T)-  \Sigma_t H^T &>& 0 \nonumber
\\
\Leftrightarrow H_r^{-1} \Sigma_d + ( H_r^{-1}  H - \mathbb{I}^{n \times n} ) \Sigma_t H^T  &>& 0 \nonumber
\\
\Leftrightarrow H H_r^{-1} \Sigma_d + (H H_r^{-1}  H - H ) \Sigma_t H^T  &>& 0 \nonumber 
\\
\Leftrightarrow \mathbb{I}^{m \times m} \Sigma_d + ( \mathbb{I}^{m \times m} H - H) \Sigma_t H^T  &>& 0 \nonumber 
\\
\Leftrightarrow \Sigma_d   &>& 0 \nonumber,
\end{eqnarray}
which holds true because $ \Sigma_d  $ is positive definite.

\section{The sNPF variance}
\label{sec:Paper3 - Appendix - sNPF variance}

The sNPF is a heuristic approach, with particle dynamics that are purely motivated by the dynamics of the first posterior moment.
As such, it is not surprising that higher order moments are, in general, not matched exactly.
Consider for instance the variance that results from the sNPF equation (Eq.~\ref{Main-eq:Paper 3 - sNPF}, main manuscript).
It is straightforward to compute by using It\^{o}'s lemma for point-process observations:
\begin{eqnarray}
d\Sigma_t & = & d\ev{\vecx_{t}\vecx_{t}^T} - d\vecmu_t \vecmu_t^T - \vecmu_t d\vecmu_t^T - (d\vecmu_t) (d\vecmu_t)^T, 
\end{eqnarray}
with 
\begin{eqnarray}
d \vecmu_t & = & \ev{\vecf_t} dt + W_t ( \vecN_{t} - \ev{\vecg_{t}} dt),
\\
d\ev{\vecx_{t}\vecx_{t}^T}  & = & \left( \ev{\vecf_t \vecx^T} + \ev{\vecx_t \vecf_t^T} + \Sigma_x \right)\, dt \nonumber \\
& & + \sum_d \left( - W_{d,t} \ev{g_{d,t} \vecx_t^T } -  \ev{\vecx_t g_{d,t} }W_{d,t}^T \right)\, dt \nonumber \\
& & + \sum_d \left( \ev{ (\vecx_t + W_{d,t}) (\vecx_t + W_{d,t})^T } - \ev{\vecx_t\vecx_t^T} \right),
\\
d \vecmu_t d\vecmu_t^T  & = & \sum_d W_{d,t} W_{d,t} ^T dN_{d,t}.
\end{eqnarray}
Thus, the evolution equation for the second moment of the sNPF reads:
\begin{eqnarray}
d\Sigma_t & = & \left( \cov(\vecx_t,\vecf_t^T) + \cov(\vecf_t,\vecx_t^T) + \Sigma_x \right) \, dt \nonumber \\
& & - \sum_d \left( W_{d,t} \cov(g_s,\vecx_t^T) + \cov( \vecx_t, g_{d,t} ) W_{d,t}^T  \right) \, dt.
\end{eqnarray}

The variance equation is completely independent of any update due to events $ dN_{d,t} $, implying that the variance does not change in response to a spike.
This result is not surprising and can be seen directly from Eq.~\eqref{Main-eq:Paper 3 - sNPF}.
Upon arrival of an event in channel $ d $, all particles jump by the same amount $ W_{d,t} $, irrespectively of their position, which leaves the absolute distance between the particles, and thus the variance estimated from the particle positions, unaltered.
From an intuitive point of view, this cannot be accurate:
since a single event conveys usually more information about the hidden state than the absence of an event, at least for this particular model we investigated here, the variance should decrease right after the spike due to increased spatial accuracy (cf.~the instantaneous update of the ADF variance in Eq.~\ref{eq:Paper 3 - Appendix - ADF dSigmaN}, which always adds a negative contribution to the variance following an event). 

We observe numerically that, for instance in the EKSPF for a nonlinear hidden dynamics, the variance tends to decrease a few iterations after the spike and on average to underestimate the true posterior variance.
If the hidden process is an OU process, the variance seems to become stationary (Figure \ref{fig:Paper3 - 1D variance linear}).
The behaviour we describe here is not directly obvious from the sNPF equation.
This could further be a possible explanation for its suboptimality.

\begin{figure}
	\centering
	\begin{subfigure}{0.49\textwidth}
		\caption{ \label{fig:Paper3 - 1D variance linear}
			Nonlinear hidden dynamics.}
		\includegraphics[width=\textwidth]{{bimodal_variance}.pdf}
	\end{subfigure}
	\begin{subfigure}{0.49\textwidth}
		\caption{Linear hidden dynamics.}
		\includegraphics[width=\textwidth]{{OU_variance}.pdf}
	\end{subfigure}
	\caption[Evolution of variance in the sNPF.]{
		\label{fig:Paper3 - 1D variance}
		{\bf Evolution of variance in the sNPF.}
		For both simulations, the same model parameters as in simulations in Figure \ref{fig:Paper3 - 1D place cell decoding - nonlinear} and \ref{fig:Paper3 - 1D place cell decoding - OU} were used.
	}
\end{figure}

\section{Particle methods as approximate solutions to the filtering problem}
\label{sec:Intro Filt - Particle methods as approximate solutions to the filtering problem}

Particle filtering is a numerical technique to approximate solutions to the filtering problem by a finite number of samples, or `particles', from the posterior.
Thus, they serve as a finite-dimensional approximation of Eq.~\eqref{eq:FS - PostExpect evolution solution}, overcoming the closure problem.
The true posterior\footnote{In the following, we use conditioning on $ \N_t $, but for the (discrete time) particle filter this is not restricted to point process observations.} is replaced by the empirical distribution formed by the particle states $ \vecx_{t}^{(i)} $, and, if it is a weighted PF, by their corresponding importance weights $ w_t^{(i)}  $,  
\begin{eqnarray}
p(\vecx_t|\N_{0:t})  & \approx& \sum_{i=1}^N w_t^{(i)} \delta ( \vecx_t - \vecx_t^{(i)} ),
\end{eqnarray}
such that 
\begin{eqnarray}
\E_\P\left[ \phi_t | \N_{0:t} \right] \approx \sum_{i=1}^N w_t^{(i)} \phi(\vecx_t^{(i)}).
\end{eqnarray}

The rationale is based on a similar idea as using the Euler-Maruyama scheme to numerically solve the Fokker-Planck equation (cf.~Section \ref{sec:Appendix - EulerMaruyama} in Appendix) and its associated equation for the posterior moments.
As a numerical recipe \cite[for instance provided by][for discrete-time models]{Doucet2000,Doucet2009}, it is easily accessible, because in principle no knowledge of the Fokker-Planck equation, nonlinear filtering theory or numerical methods for solving partial differential equations is needed. 

In this section, weighted particle filters will be introduced from a continuous-time perspective based on the change of probability measure formalism \cite[roughly following][Chapt.~9.1, and extending this to point-process observations]{Bain2009}.
From this formalism, we derive dynamics for the weights, which is the analytical basis for our analysis of the COD in Section \ref{Main-sec:Paper3 - COD} in the main manuscript.
Finally, to give context for readers more familiar with discrete-time particle filtering, the continuous-time perspective will be linked to the `standard' particle filter (PF).

\subsection{Particle filtering in discrete time }
\label{sec:Intro Filt - Particle filtering in discrete time}

Here, we will briefly outline the discrete-time algorithm, following \cite{Doucet2000}, and clarify how we used it with our generative model.\footnote{Modified from \cite[SI, Section S1.3.1]{Kutschireiter2017}}

In the discrete-time filtering literature (cf.~for instance \cite{Doucet2009}), one considers a state-space model of the form
\begin{eqnarray}
\vecx_t & \sim & p(\vecx_t|\vecx_{t-1} ), \label{eq:Intro Filt - PF prior transition} \\
\vecy_t & \sim  & p(\vecy_t|\vecx_{t}), \label{eq:Intro Filt - PF emission}
\end{eqnarray}
where $ \vecx_t $ denotes the hidden process following the prior transition probability $ p(\vecx_t|\vecx_{t-1} ), $ and $ \Y_{0:t} = \{ \vecy_0,\dots , \vecy_t \} $ is the sequence of observations generated from the emission probability $ p(\vecy_t|\vecx_{t}) $ at each time step.
For our model, the observations correspond to the increments of the point process in Eq.~\eqref{eq:GM - dN}, such that $ p(\vecy_t|\vecx_{t}) = p(dN_t|\vecx_{t}) $.

In sequential importance sampling, sampling and reweighing is done recursively at each time step $ t $
\begin{eqnarray}
\vecx_t^{(i)} & \sim & \pi (\vecx_t | \X_{0:t-1}^{(i)}, \Y_{0:t}), \label{eq:Intro Filt - PF proposal} \\
\tilde{w}_t^{(i)} & = & \tilde{w}_{t-1}^{(i)} \frac{p(\vecy_t|\vecx_t^{(i)})\, p(\vecx_t^{(i)}|\vecx_{t-1}^{(i)} ) }{\pi (\vecx_t^{(i)} | \X_{0:t-1}^{(i)}, \Y_{0:t})}, \label{eq:Intro Filt - PF weights unnormalized}\\
w_t^{(i)} & = & \frac{\tilde{w}_t^{(i)}}{\sum_{j=1}^{N} \tilde{w}_t^{(j)} }, \label{eq:Intro Filt - PF weights}
\end{eqnarray}
where $ \pi (\vecx_t | \X_{0:t-1}^{(i)}, \Y_{0:t}) $ denotes the proposal function, and $ \tilde{w}_t^{(i)} $ denotes the unnormalized importance weight of particle $ i $ at time $ t $.

A convenient choice is to use the prior transition probability in Eq.~\eqref{eq:Intro Filt - PF prior transition} as the proposal function in Eq.~\eqref{eq:Intro Filt - PF proposal}.
Then, computation of the unnormalized weights simplifies to
\begin{eqnarray}
\tilde{w}_t^{(i)} & = & \tilde{w}_{t-1}^{(i)} p(\vecy_t|\vecx_t^{(i)}). \label{eq:Intro Filt - bootstrap weights unnormalized}
\end{eqnarray}
This scheme is the basis of the famous \textbf{Bootstrap PF} (BPF \cite{Gordon1993})\footnote{Although technically, the BPF requires a resampling step at every iteration step.}.
Doucet et al.~\cite{Doucet2000} state that the BPF is ``inefficient in simulations as the state-space is explored without any knowledge of the observations'', which is certainly true.
However, the importance of this `simple' choice seems to be underestimated:
for instance, it turns out that in the continuous-time limit, the 'optimal proposal function' proposed in the same reference, which is supposed to take into account the observations in such a way that the expected variance of the importance weights is minimized, becomes the prior transition probability (cf.~Appendix of \cite{Surace2016}).

\subsection{Particle filtering in continuous time}
\label{sec:Intro Filt - Particle filtering in continuous time}

Based on sequential importance sampling, both samples (or 'particles') $ \vecx_t^{(i)} $ as well as their respective weights are propagated through time.
For continuous time problems, importance sampling amounts again to a change of measure from the original measure $ \P $ to a reference measure $ \Q $ under which the observation processes are independent of the hidden process.

Why this should be the case is rather intuitive when recalling the Kallianpur-Striebel formula \eqref{eq:FS - KallianpurStriebel}:
\begin{eqnarray}
\E_\P\left[ \phi_t | \N_t \right] & =& \frac{1}{Z_t} \E_\Q [\phi_t\, L_t | \N_t ]. \nonumber
\end{eqnarray}
If we want to approximate the left-hand side of this equation with empirical samples, it would require us to have access to samples from the real posterior distribution, which is usually not the case.
However, since under the measure $ \Q $ on the right-hand side hidden state and observations are decoupled, the estimate is approximated by empirical samples that correspond to realizations of the hidden process (Eq.~\ref{eq:GM - hidden}):
\begin{eqnarray}
\frac{1}{Z_t} \E_\Q [\phi_t\, L_t | \N_t ] & \approx & \frac{1}{\bar{Z}_t} \sum_{i=1}^{P} \phi(\vecx_t^{(i)}) L_t (\vecx_t^{(i)}).
\end{eqnarray}
$ \bar{Z}_t = \sum_{i=1}^P L_t (\vecx_t^{(i)}) $ is an empirical estimate of the normalization constant, and the Radon Nikodym derivative $ L_t $ is given by Eq.~\eqref{eq:FS - RadonNikodym}.

Evaluating the Radon Nikodym derivative at the particle states $ \vecx_{t}^{(i)} $, the importance weight $ w_t^{(i)} $ of particle $ (i) $ at time $ t $ for point-process observations in continuous time is given by:
\begin{eqnarray}
w_t^{(i)} & = &  \frac{1}{\bar{Z}_t} \prod_{d = 0}^m  \exp \left( \int_{0}^{t} \log\frac{g_d(\vecx_{s}^{(i)})}{\lambda_0} dN_{d,s} +  \int_{0}^{t} (\lambda_0 - g_d (\vecx_s^{(i)})  ) \, ds \right). \label{eq:Intro Filt - cont time  weights PP}
\end{eqnarray}

\subsubsection{Weight dynamics in continuous time}
\label{sec:Weight dynamics in continuous time}

We are interested understanding in how the weight of particle $ (i) $ changes over time, and therefore derive an evolution equation for the normalized importance weights.
\begin{eqnarray}
d w_t^{(i)} & = & d \left( \frac{L_t(\vecx_t^{(i)})}{\bar{Z}_t}  \right) \nonumber \\
&= & \bar{Z}_t^{-1} dL_t^{(i)} + L_t^{(i)}d\bar{Z}_t^{-1} + dL_t^{(i)} \, d\bar{Z}_t^{-1}, \label{eq:Intro Filt - weight dynamics general}
\end{eqnarray}

With Eq.~\eqref{eq:FS - RadonNikodym evolution} we find for the dynamics of the weights:
\begin{eqnarray}
dL_t^{(i)} & = & L_t^{(i)} \sum_{d=0}^{m} \frac{1}{\lambda_0} \left( g_d (\vecx_t^{(i)}) - \lambda_0 \right) \left(dN_{d,t} - \lambda_0 \, dt  \right), \\
d\bar{Z}_t &=& \bar{Z}_t  \sum_d  \frac{1}{\lambda_0} (\bar{g}_d - \lambda_0) \left(  dN_{d,t} - \lambda_0 \, dt \right),
\end{eqnarray}
and thus, using It\^{o}'s lemma for point processes to obtain $ d\bar{Z}_t^{-1} $, with Eq.~\eqref{eq:Intro Filt - weight dynamics general}:
\begin{eqnarray}
dw_t^{(i)} & = & w_t^{(i)} \sum_d \frac{1}{\bar{g}_{d,t}} \left( g_d(\vecx_t^{(i)}) - \bar{g}_{d,t} \right) \left(  dN_{t,d} - \bar{g}_{d,t} dt \right), \label{eq:Intro Filt - weight dynamics PP}
\end{eqnarray}
with $ \bar{g}_{d,t} = \sum_{i=1}^P g_d(\vecx_t^{(i)})  $.

\subsection{Effective number of particles and `curse of dimensionality'}
\label{sec:Intro Filt - Effective number of particles and curse of dimensionality}

Particle filtering approaches are known to suffer from an inevitable weight decay over time \cite{Doucet2000}, i.e.~without numerical intervention such as resampling, after a few iterations the particle weights will be highly degenerate, i.e.~only a few (in very extreme cases only one) particle carry significant weights, whereas the rest of the weights are almost equal zero.
Thus, the posterior density is effectively approximated by much less particles than the actual number of particles in the system, which can lead to severe deficiency in filtering performance.
Further, it is computationally disadvantageous, because one has to propagate a lot of particles that do not contribute to the posterior estimate. 
In \cite{Surace2016} the problem of weight decay is linked to the so-called `curse of dimensionality' (COD), which manifests itself in an exponential growth in the number of particles that are required to maintain a certain numerical performance.

Our analytical argument in Section \ref{Main-sec:Paper3 - COD} in the main manuscript relies on the effective number of particles, which is an empirical measure that quantifies the level of weight degeneracy:
\begin{eqnarray}
P_\text{eff} & = & \frac{1}{\sum_{i=1}^P (w_t^{(i)})^2},
\end{eqnarray}
with $ 1 \leq P_\text{eff} \leq P $
The effective number of particles approximately estimates the equivalent of independent samples drawn directly from the posterior distribution \cite{Martino2017}.

For an estimate of the time scale on which the effective number of particles depletes, it suffices to analyse the dynamics of the \emph{inverse} of $ P_\text{eff} $, which can be obtained by using It\^{o}'s lemma together with Eq.~\eqref{eq:Intro Filt - weight dynamics PP}.
Thus, we find
\begin{eqnarray}
d P_\text{eff}^{-1} & = &  \sum_{i=1}^P d (w_t^{(i)})^2  \nonumber 
\\
& = & \sum_{i=1}^P(w_t^{(i)})^2 \sum_{d=1}^{m}\left[ -2 (g_{d,t}^{(i)} - \bar{g}_{d,t})\,dt + ((g_{d,t}^{(i)})^2 -\bar{g}_{d,t}^2) \frac{1}{\bar{g}_{d,t}^2} \, dN_{d,t} \right]. \label{eq:Intro Filt - dPeff^-1 PP}
\end{eqnarray}

It is not straightforward to estimate the depletion time scale based on this equation.
However, for a general generative model, Eq.\eqref{eq:Intro Filt - dPeff^-1 PP}, still allows for a `back of the envelope' estimate:
if we consider initialization with $\w{i}_0=\frac{1}{P} $ $ \forall i $, then $ P_\text{eff}^{-1} $ can only grow.
Since its dynamics depends on a sum with $ m $ summands, it is proportional to the number of observable dimensions $ m $.
Thus, the initial growth rate of $ P_\text{eff}^{-1} $ is estimated to be proportional to $ 1/m $ and in turn the decay rate of $ P_\text{eff} $ is proportional to $ m $.
We would like to add that numerical simulations (\ref{Main-fig:Paper 3 - COD} in main manuscript) showed that this estimate is in practice inaccurate.
In fact, it seems that the decay is even faster than our estimate, indicating that the point process $ \N_t $ probably adds to the estimate in a nontrivial way.

\section{Maximum Likelihood Parameter learning}

\subsection{The likelihood function}
\label{sec:The likelihood function}

Consider being given a realization (or trajectory)  $ \mathcal{Y}_t = \{ y_0, \dots, y_t \} $ of a stochastic process, which may have been generated by two different models of the stochastic process, which induce two different probability measures $ \P $ and $ \Q $, respectively.
Then the Radon-Nikodym derivative between the two probability measures evaluated at $ \mathcal{Y}_t $, $ L_t (\mathcal{Y}_t) = \frac{dP}{dQ} (\mathcal{Y}_t ) $, determines which model is more likely to have generated the trajectory.
Therefore, the Radon-Nikodym derivative is also referred to as likelihood function \cite{Klebaner2005}, and its logarithm as the log likelihood function, which we denote by $ \Lambda_t $.

In the filtering context, we consider a the Radon-Nikodym derivative between the original measure $ P_\theta $ and a reference measure $ \Q $, which is independent of the model parameters, and maximize it with respect to the parameters $ \theta $.
Conveniently, choosing $ \Q $ to be the measure under which the observations follow a Poisson process with a constant rate $ \lambda_0 $, the observations are independent of both the hidden state process $ \vecx_t $ as well as the model parameters $ \theta $.
However, we are not given a the whole realization of the stochastic process, i.e. $ (\mathcal{X}_t,\N_{0:t}) = \left\{ (\vecx_0,\vecN_0),\dots, (\vecx_t,\vecN_t) \right\}  $, but only partial information in terms of a sequence of observations $ \N_{0:t} $.
Therefore, we need to marginalize over the hidden state process\footnote{Formally, we restrict the measures $ \P $ and $ \Q $ to the filtration $ \mathcal{F}_t^\mathcal{N} $ and then compute the Radon-Nikodym derivative of the restricted measures, which can be expressed as conditional expectation under $ \Q $.}:
\begin{eqnarray}
\Lambda_t (\N_{0:t},\theta) & = & \log \E_\Q \left[\frac{dP_\theta}{dQ} | \N_{0:t} \right] = \log \E_\Q \left[ L_t^\theta | \N_{0:t} \right],
\end{eqnarray}
where the Radon-Nikodym derivative $ L_t^\theta $ is given by Eq.~\eqref{eq:FS - RadonNikodym}.

By making use of Eq.~\eqref{eq:FS - RadonNikodym evolution} and the fact, that under the measure $ \Q $ the stochastic differential and the expectation can be interchanged, we can compute the expectation on the right hand side, and consequently derive an expression for the likelihood $ \Lambda_t $.
\begin{eqnarray}
d \E_\Q \left[ L_t^\theta | \N_{0:t} \right] & = &  d\expect{L_t^\theta}_\Q =  \expect{dL_t^\theta}_\Q \nonumber
\\
& = & \expect{ L_t^\theta \cdot \sum_{d = 1}^{m} \left( \frac{g_d^\theta(\vecx_t)}{\lambda_0}  -1 \right) (dN_{d,t} - \lambda_0 \, dt)}_\Q \nonumber
\\
& = & \sum_{d = 1}^{m} \left( \frac{ \expect{L_t^\theta \cdot g_d^\theta(\vecx_t)}_\Q }{\lambda_0}  - \expect{L_t^\theta}_\Q  \right) (dN_{d,t} - \lambda_0 \, dt) \nonumber
\\
& = &\expect{L_t^\theta}_\Q  \sum_{d = 1}^{m} \left( \frac{ \expect{ g_d^\theta(\vecx_t) }_{P_\theta} }{\lambda_0}  - 1 \right) (dN_{d,t} - \lambda_0 \, dt). 
\end{eqnarray}
Noting the structural similarity of this SDE with Eq.~\eqref{eq:FS - RadonNikodym evolution}, the solution is structurally similar to Eq.~\eqref{eq:FS - RadonNikodym}, but now the posterior expectation $  \expect{ g_d^\theta(\vecx_t) }_{P_\theta} $ is used instead of $ g_d(\vecx_t) $.
Consecutively taking the logarithm, we arrive at the log likelihood function:
\begin{eqnarray}
\Lambda_t(\N_{0:t},\theta) & = & \log \left[  \prod_{d = 0}^m \left( \prod_{0 <s \leq t} \frac{\expect{ g_d^\theta(\vecx_t) }_{P_\theta} }{\lambda_0} \Delta N_{d,s}   \right)  \exp \left( \int_{0}^{t} (\lambda_0 - \expect{ g_d^\theta(\vecx_t) }_{P_\theta}   ) \, ds \right) \right]  \nonumber
\\
& = & \int_{0}^{t} \sum_{d=0}^{m} \left[ (\lambda_0 - \expect{ g_d^\theta(\vecx_t) }_{P_\theta} )\, ds + \log\frac{\expect{ g_d^\theta(\vecx_t) }_{P_\theta} }{\lambda_0} \, dN_{d,s} \right]. \label{eq:PL - log likelihood}
\end{eqnarray}
Notably, this log likelihood function is expressed in terms of the parametrized filter, and thus allows us to be computed online from the filter output.

\subsection{Filter derivative dynamics}
\label{sec:Paper3 - Appendix - Filter derivatives and learning rules}

The learning rules in Eq.~\eqref{eq:Paper3 - learning rules offline} and \eqref{eq:Paper3 - learning rules online} depend on the derivative of the posterior expectation of the rate function.
Using the sNPF, we approximate this expectation with equally weighted particles according to Eq.~\eqref{eq:Paper 3 - sNPF particle representation}:
\begin{eqnarray}
\ev{\vecg_{t}} & = & \frac{1}{N} \sum_{i=1}^P \vecg (\vecx_t^{(i)}),
\end{eqnarray}
such that
\begin{eqnarray}
\partial_\theta \ev{\vecg (\vecx_t) } & = & \frac{1}{P} \sum_{i=1}^P \left(  \partial_\theta \vecg(\vecx_t) |_{\vecx_t=\vecx_t^{(i)}} + \left( D\vecg_t\cdot \partial_\theta \vecx \right) |_{\vecx_t=\vecx_t^{(i)}} \right) \nonumber
\\
& = & \ev{\partial_\theta \vecg_{t}} + \ev{ D\vecg_t \cdot \vecalph_{\theta,t}}, \label{eq:Paper 3 - Appendix - Derivative ev g}
\end{eqnarray}
where $ \left( D\vecg_t \right)_{i,j} = \partial_{x_j} g_i$ denotes the Jacobian of the function $ \vecg_t $.
Here, we defined the \textbf{particle filter derivative} $ \vecalph_{\theta,t}^{(i)} :=  \partial_\theta \vecx_t^{(i)} $.

The dynamics of the particle filter derivative is computed from the particle dynamics in Eq.~\eqref{eq:Paper 3 - sNPF}, and evolved and integrated alongside the sNPF.
It reads:
\begin{eqnarray}
d \vecalph_\theta^{(i)} & = & \partial_\theta d \vecx_t^{(i)}  \nonumber
\\
& = & \partial_\theta \vecf_t^{(i)} \, dt + \partial_\theta \Sigma_x d \vecw_t^{(i)} \nonumber \\
& & + (\partial_\theta  W_t ) \left( d\mathbf{N}_t - \mathbf{g}_t^{(i)} \,dt \right) -  W_t \,  \partial_\theta \vecg_t^{(i)}\,dt, 
\end{eqnarray}
with
\begin{eqnarray}
\partial_\theta \vecf^{(i)}_t  & = & \partial_\theta \vecf(\vecx_t) |_{\vecx_t=\vecx_t^{(i)}} + \left( D\vecf_t\cdot \partial_\theta \vecx \right) |_{\vecx_t=\vecx_t^{(i)}} .
\end{eqnarray}

For the EKSPF, the $ l $-th column of the gain matrix $ W^{(l)} $ is determined according to $ W^{(l)} = \frac{\cov(g^{(l)} , \vecx_t)}{\expect{g^{(l)}}}  $.
If we want to use this filter for learning, we need compute partial (and filter) derivatives of the gain as well.
The derivative for the gain then reads:
\begin{eqnarray}
\partial_\theta W_{l,t} & = & \frac{1}{\ev{g_{l,t}}} \left( \cov( \partial_\theta g_{l,t} , \vecx_t) + \cov( \nabla_\vecx^T g_l \cdot \vecalph_{\theta,t} , \vecx_t) + \cov( g_{l,t} , \vecalph_{\theta,t} )  \right) \nonumber
\\
& & - \frac{1}{ \ev{ g_{l,t} }^2 } \ev{\partial_\theta g_{l,t}} \cov(g^{(l)} , \vecx_t).
\end{eqnarray}

\section{Simulation of the RSDE}
\label{sec:Paper3 - Appendix - Details on numerical experiments}

In Eq.~\eqref{eq:Paper 3 - 2d model dx}-\eqref{eq:Paper 3 - 2d model - boundary process}, we model the movement of an animal in a confined box as an RSDE.
Numerically, there are several ways to implement such a process (cf.~\cite{Brillinger2003} for a quick overview), and here we will focus on the projection method (cf.~\cite{Lepingle1995,Hanks2017}).
Briefly, this implies taking a sample from the \emph{unconstrained} process $ \tilde{\vecx}_t $, which, should if fall outside the region $ \D $, is projected back into $ \D $:
\begin{eqnarray}
\vecx_{t+1} & = & \Pi_\D ( \tilde{\vecx}_{t+1} ) = \Pi_\D (\vecx_{t} + \vecv_t \, \delta t ),
\end{eqnarray}
which practically implies
\begin{eqnarray}
\vecx_{t+1} & = & \text{argmin}_{\mathbf{u} \in \D} \left( ||\mathbf{u} - \tilde{\vecx}_{t+1} \right).
\end{eqnarray}
Further, since this projection implies that there is no movement perpendicular to the boundary, upon projection the velocity component has to be updated accordingly (i.e.~the component perpendicular to the boundary has to be zero after projection).

When implementing a realization from this process with boundary conditions, one should keep in mind that a standard Euler-Maruyama scheme does not have the same rate of convergence as would be the case without boundary conditions.
\cite{Lepingle1995} gives an intuitive explanation why this might happen.
After one simulation step with a boundary crossing, the same final \emph{unprojected} state could have been reached by crossing the border not once, but several times due to stochastic movement, which in turn would lead to significantly different outcomes for the trajectory.
In our model, this is fortunately not so severe, since the stochasticity only enters through the velocity components, which are generally unconstrained.

\bibliographystyle{plain}
\bibliography{../Bibliography/library}